\documentclass[fleqn,10pt]{olplainarticle}
\usepackage{timet}

\usepackage{amsmath}
\usepackage{amsfonts}
\usepackage{graphicx}
\usepackage{physics}
\usepackage{amsmath}
\usepackage{amsfonts}
\usepackage{graphicx}
\usepackage{physics}
\usepackage{fancyhdr}
\usepackage{hyperref}

\newcommand\widthFig{0.42}

\title{Flexible quasi-2D inversion of time-domain AEM data, using a wavelet-based complexity measure}
\author[Wouter Deleersnyder]
{Wouter Deleersnyder$^{1,2}$, Benjamin Maveau$^{1}$, Thomas Hermans$^{2}$, David Dudal$^{1,3}$\\
	$^{1}$KU Leuven Campus Kortrijk - KULAK, Department of Physics, Etienne Sabbelaan 53, 8500 Kortrijk, Belgium.\\  E-mail: {wouter.deleersnyder@kuleuven.be} \\
	$^{2}$Ghent University, Department of Geology, Krijgslaan 281 - S8, 9000 Gent, Belgium \\
	$^{3}$Ghent University, Department of Physics and Astronomy, Ghent, Krijgslaan 281 - S9, 9000 Gent, Belgium \\
}
\date{Accepted 2023 January 20. Received 2022 December 23; in original form 2022 October 3}


\begin{abstract}
	Regularization methods improve the stability of ill-posed inverse problems by introducing some a priori characteristics for the solution such as smoothness or sharpness. In this contribution, we propose a multidimensional, scale-dependent wavelet-based $\ell_1$-regularization term to cure the ill-posedness of the airborne (time-domain) electromagnetic induction inverse problem. The regularization term is flexible, as it can recover blocky, smooth and tunable in-between inversion models, based on a suitable wavelet basis function. For each orientation, a different wavelet basis function can be used, introducing an additional relative regularization parameter. We propose a calibration method to determine (an educated initial guess for) this relative regularization parameter, which reduces the need to optimize for this parameter, and, consequently, the overall computation time is under control.
	We apply our novel scheme to a time-domain airborne electromagnetic data set in Belgian saltwater intrusion context, but the scheme could equally apply to any other 2D or 3D geophysical inverse problem. 
\end{abstract}

\begin{document}
	
	\flushbottom
	
	\maketitle
	\textbf{\color{red} Published in Geophysical Journal International} \url{https://doi.org/10.1093/gji/ggad032}\\
	\thispagestyle{empty}

\section{Introduction}

The Airborne ElectroMagnetic induction (AEM) method is a practical tool to map near-surface geological features over large areas (a few tens of kilometres) via the bulk electrical resistivity. It is increasingly used for mineral exploration \citep{macnae2007developments}, hydrogeological mapping \citep{mikucki2015deep, podgorski2013processing}, saltwater intrusion \citep{goebel2019mapping, siemon2019automatic} and contamination \citep{pfaffhuber2017delineating}. The focus in AEM is mainly on (dual moment) time-domain systems, as they allow for an improved near-surface resolution and an increased depth of penetration. While the AEM systems have massively advanced within the last decades \citep{auken2017review}, there are two main impediments in the data interpretation process, which are also a concern for other electromagnetic geophysical methods. The first difficulty is the computational burden related to the computation of the forward model, which describes the subsurface response to a specific subsurface realization and specific survey set-up. To date, most inversion schemes use a one dimensional forward model, which assume horizontal layers without lateral variations. The second challenge lies in the fact that geophysical inversion (in general) is an ill-posed problem, meaning that the solution is typically not unique. This is usually dealt with via regularization techniques.

Deterministic regularization techniques impose constraints on the model parameters. It is generally a minimum-structure inversion, meaning that unrealistic electrical conductivities are filtered out and the ``simplest'' inversion model is promoted. ``Simple'' means that we aim for the model with the least number of features that explains the data equally well as other more complex models with more (potentially attractive) features. This is Occam’s inversion approach articulated by \citet{constable1987occam}, who popularised smoothness based regularization for electromagnetic sounding data. The smoothness constraint, such as the traditional Tikhonov regularization \citep{tikhonov1943stability}, is not always adapted to the subsurface structure \citep{linde2015geological}. Smoothness regularization improves the stability of the inversion, however, too much regularization smears out small-scale features, while little regularization reduces the stability. Moreover, blocky structures or sharp interfaces cannot be recovered. While many alternative minimum-structure schemes exist, it is not always easy to find the best adapted scheme. One flexible inversion scheme supporting multiple types of minimum-structure constraints could simplify the generation of an ensemble of various inversion models.

Quasi-2D inversion schemes produce realistic 2D inversion models by imposing regularization conditions in both dimensions. In contrast to 2D inversion schemes, these schemes rely on 1D forward modelling due to the computational burden of accurate 2.5D or 3D modelling. Common 2D regularization schemes use smoothing, minimization of the total variation or wavelet theory. A traditional method for minimum structure inversion is to apply a smoothing constraint, such as with Tikhonov regularization along all the orientations \citep{tikhonov1943stability}.  Tikhonov regularization promotes smooth solutions and hence cannot recover blocky structures. Laterally Constrained Inversion (LCI) \citep{auken2004layered, siemon2009laterally} is capable of producing laterally smooth transitions by adding roughening constraints to the objective function that tie model parameters of adjacent layers. Via a roughening matrix, the dissimilarity between neighbouring cells is measured and minimized in combination with a data misfit. Amongst its successors are Spatially Constrained Inversion (SCI) for smooth quasi-3D inversion by \citet{viezzoli2008quasi} and sharp SCI, which favours more blocky models \citep{vignoli2015sharp, klose2022laterally}, using a minimum gradient support functional. Interestingly, the latter work allows for a tunable bi-directional sharpness/smoothness. Another similar approach is the lateral parameter correlation \citep{christensen2016strictly}. This approach firstly inverts the data without constraints. Then, a laterally smooth version of the inversion model is generated. The method finishes with a final inversion on a starting model which is a result of a covariance analysis of the constrained and unconstrained inversion model. Total variation regularization methods with several focusing functions are mostly sparsity-based, such as the $\ell_1$-norm, and successfully recover blocky structures \citep{farquharson2007constructing}. Other examples use the minimum gradient support functional in a larger workflow, such as \citet{thibaut2021new}, or provide methods to tune the parameter in the minimum gradient support functional \citep{deleersnyder2022determining}.. There are many alternatives such as covariance-based inversion \citep{hermans2012imaging, paasche2007cooperative}. Recently, (sparsity-based) wavelet-based regularization schemes have been applied to geophysical inversion. \cite{nittinger2016inversion} present a 2D, wavelet-based sparsity inversion scheme for magnetotelluric data based on \cite{daubechies2004iterative}’s work on the iterative soft thresholding algorithm for solving objective functions with both an $\ell_1$ and $\ell_2$-norm. A particular complex dual-tree wavelet representation, yielding six directions of a smooth shape, is utilised. \citet{liu2017wavelet} apply a 3D wavelet-based method to frequency AEM data and compare the effect of various wavelet basis functions to the inversion result. \cite{liu2017wavelet} conclude that smooth wavelet basis functions produce more stable results. \citet{su2021sparse} use the shearlet transform to add (multi)directionality into the inversion model. \cite{nittinger2018compressive} already report an advantageous feature of wavelet-based inversion, that of the simultaneous occurrence of smooth and sharp anomalies within the same model. In our earlier work \citep{deleersnyder2021inversion}, this observation is confirmed with a novel scale-dependent wavelet-based regularization scheme for 1D geophysical inversion. The scale-dependency produces stable inversion models for all the common wavelet basis functions (also the blocky and irregular-shaped basis functions), which leads to a more flexible inversion scheme (as it can recover both blocky, intermediate and smooth profiles). All the 2D inversion schemes listed above are most successful with the inversion of either blocky or smooth results, but fail to generate an ensemble of inversion models with different features with the ease of changing a simple tuning parameter. In this work, the flexible scheme of \cite{deleersnyder2021inversion} is extended into two dimensions. The generalization of the scale-dependent scheme comes with a few challenges and design preferences, which are described below.\\

The regularization scheme in this work differs from the other 2/3D wavelet-based methods first and foremost because of the implementation of scale-dependency of the complexity measure proposed in \citet{deleersnyder2021inversion} for 1D inversion. Additionally, we consider a complexity measure per orientation. This allows imposing a different type of structure (blocky, smooth, ...) on the inversion model for each orientation. Due to this design choice, we can no longer optimize the inverse problem in the wavelet domain (as in our earlier work). First, we would have (at least) twice as many parameters to optimize as in model space, which makes the inverse problem more computationally expensive. Secondly, the recovered inversion model in the wavelet domain would be difficult to consistently back transform to the model space. Therefore, in this work, the inverse problem is solved in model space instead, by making use of the chain rule.

The separation of the complexity measure into an individual complexity measure for each orientation introduces an additional relative weighting parameter. To tackle this, we introduce a calibration step that we can deploy prior to the 2D inversion and yield an estimate for that relative weighting parameter. This data-driven calibration step is based on heuristics and uses the variation in the EM field data as an estimate for the lateral complexity. This calibration step is further described in Section \ref{sec:calibration}. 

In Section \ref{sec:results}, a synthetic model and field data case in a salinization context is presented. The field AEM data is from Flanders Environment Agency \citep{vlaanderentopsoil} and is the basis of Flanders' current salinization map, showing the depth of the fresh-saltwater interface.

\section{Methods}
\subsection{The objective function}
Due to the non-linearity of accurate time-domain electromagnetic (TDEM) forward operators, the inverse problem is solved iteratively as an optimization problem, where an objective function is minimized. The objective function $\phi$ takes the model parameters $\mathbf{M}$ (the electrical conductivities)  as input and expresses how well those parameters fit the data. In minimum structure inversion procedures, an additional measure of model complexity (also known as regularization or model misfit term) is added to the objective function in order to stabilize the inverse problem. The minimum of the objective function $\phi$ is usually obtained via gradient based methods. Here we use the L-BFGS-B optimization method \citep{zhu1997algorithm}, as implemented in SciPy \citep{scipy}.\\ 

In general, the objective function for a geophysical inversion problem is
\begin{equation}
	\phi(\vb{M}) = \phi_d + \beta\phi_m,
\end{equation}
where $\phi_d$ and $\phi_m$ are the data misfit and model misfit, respectively. $\beta$ is a regularization parameter which balances the relative importance of the two misfits. When the regularization parameter $\beta$ is too small, the optimization algorithm will be over-fitting the data and the geological interpretation of the inversion model is problematic. Too large regularization parameters yield too simple inversion models.\\

The data misfit functional $\phi_d$ measures how well the model parameters $\vb{M}$ fit the data. In our inversion scheme, we use the traditional weighted least-squares data-fitting term or chi-square misfit function
\begin{equation}
	\phi_d = \frac{1}{n_d}||\vb{W}_d\left(\vb{d} - \mathcal{F}(\vb{M})\right)||^2_2,
\end{equation}
where $n_d$ is the number of data points, the vector $\vb{d} \in \mathbb{R}^{n_d}$ contains the observed data of the AEM survey and $\mathcal{F}(\vb{M})$ is the predicted data via the forward model $\mathcal{F}$ (see Section \ref{sec:forward}). The diagonal matrix $\vb{W}_d$ contains the reciprocals of the estimated noise standard deviation. As AEM data span multiple orders of magnitude, the percent uncertainty is especially important, it prevents over-fitting the large values at the expense of under-fitting the small data values. From statistical theory, the chi-square criterion can be adopted \citep{kemna2000tomographic}, meaning that a good inversion model fits the data within the noise level and that the data misfit $\phi_d$ (or noise weighted root mean-squared error $\epsilon_{\text{RMS}} = \sqrt{\phi_d}$) is close to 1. This is the discrepancy principle \citep{hansen2010discrete} and will be combined with a $\beta$-cooling strategy to reduce the computational burden. The specific details are described in Appendix \ref{ap:betareducing}.\\

The model misfit functional $\phi_m$ imposes the additional constraints on the inversion model in an Occam's sense and handles the ill-posedness of the problem. In contrast to parametric models, voxel-based models typically contain much more model parameters and consequently the solution is not unique. 

\begin{figure*}
	\begin{tabular}{@{}l@{}}
					A. \hspace{\widthFig\linewidth} B. \\
			\includegraphics[width=\widthFig\textwidth]{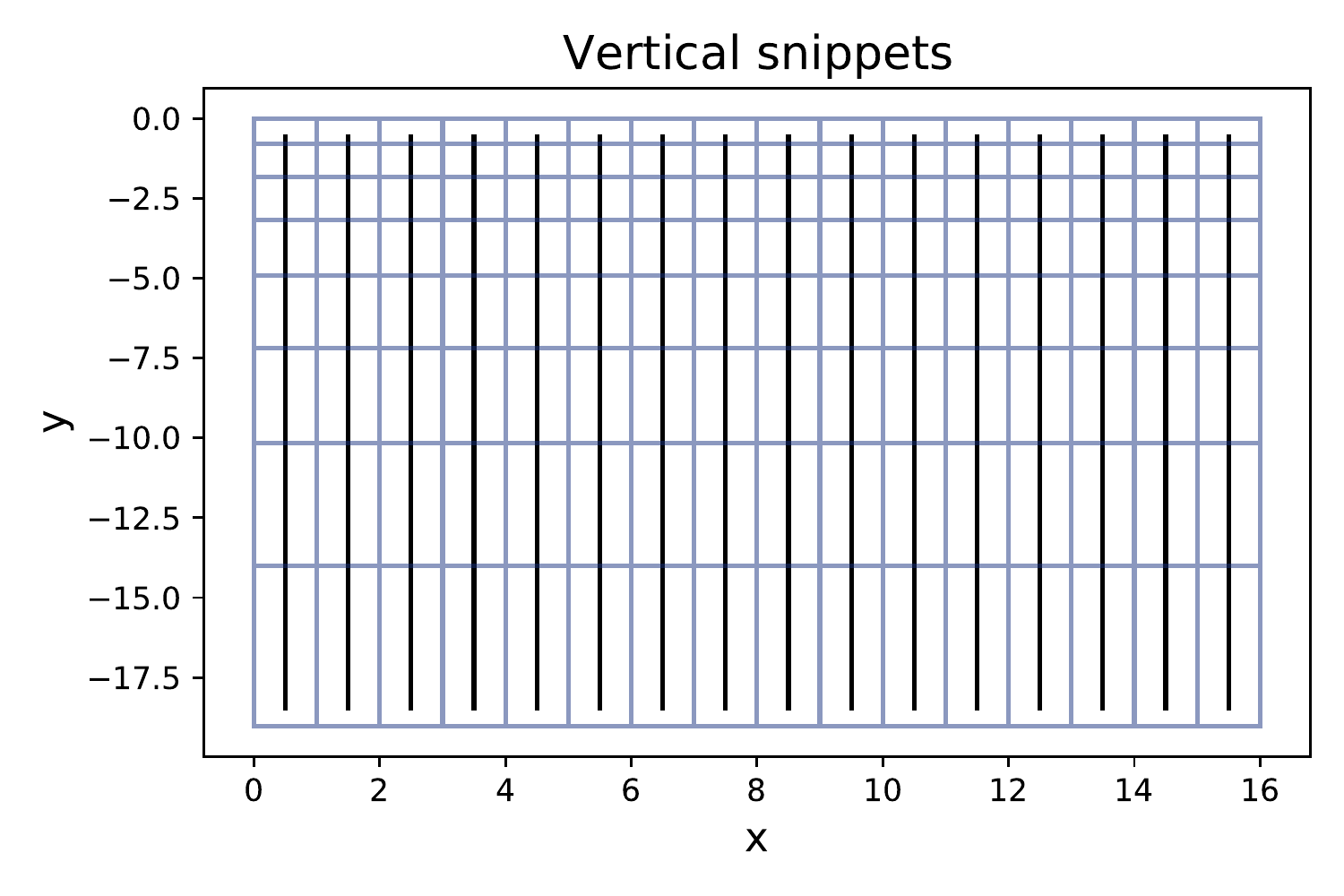}  \includegraphics[width=\widthFig\textwidth]{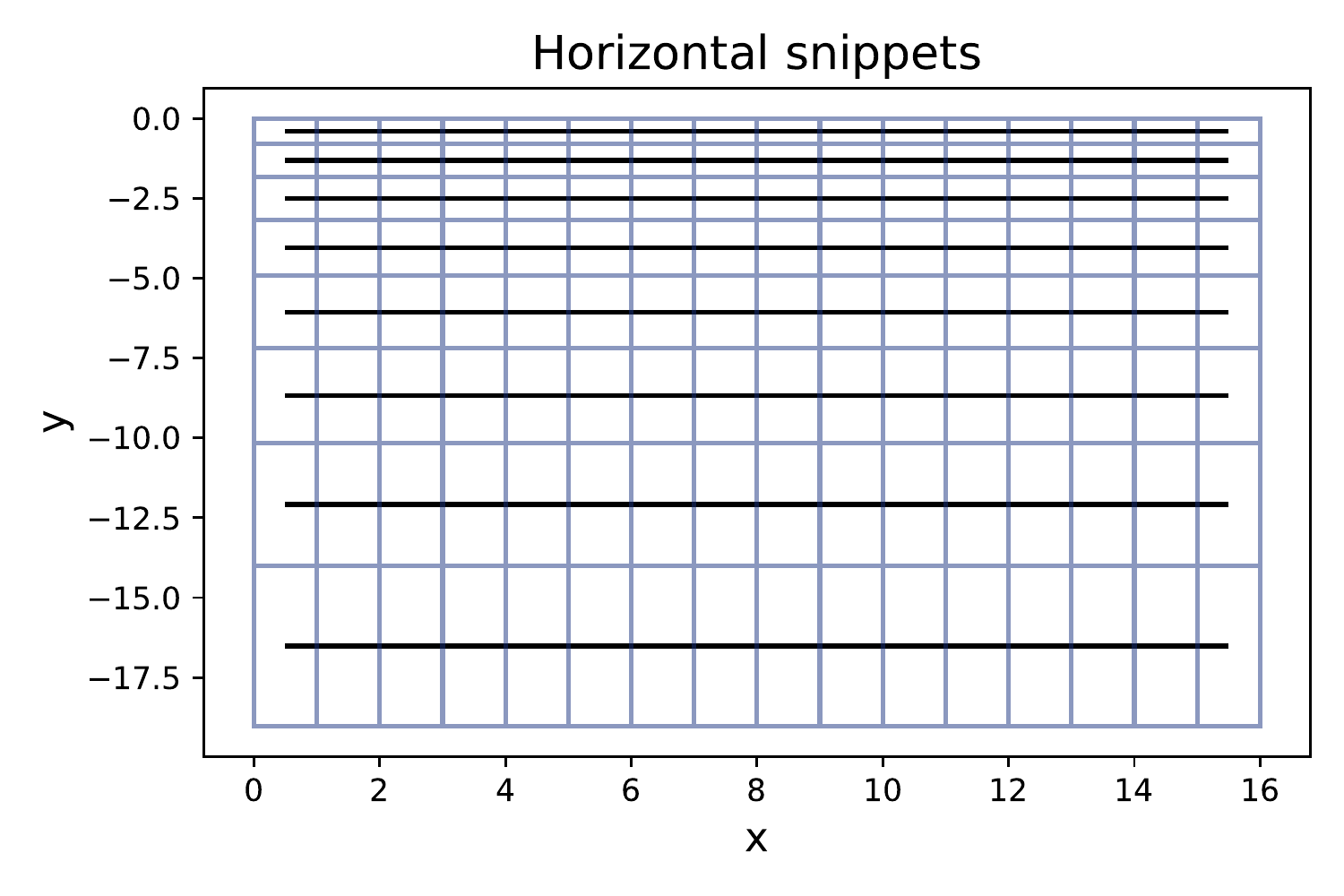}  \\
					C. \hspace{\widthFig\linewidth} D. \\
			\includegraphics[width=\widthFig\textwidth]{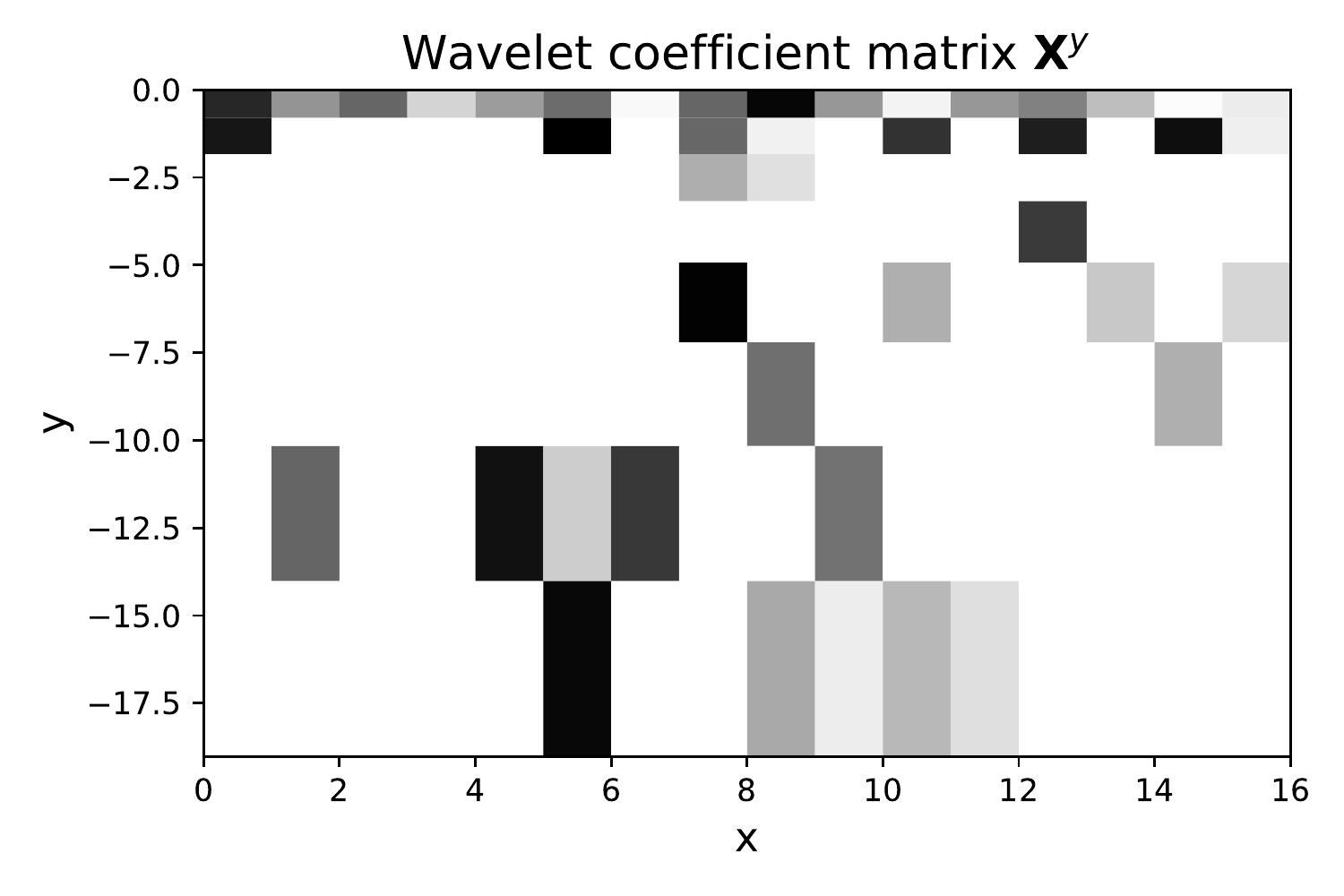}  \includegraphics[width=\widthFig\textwidth]{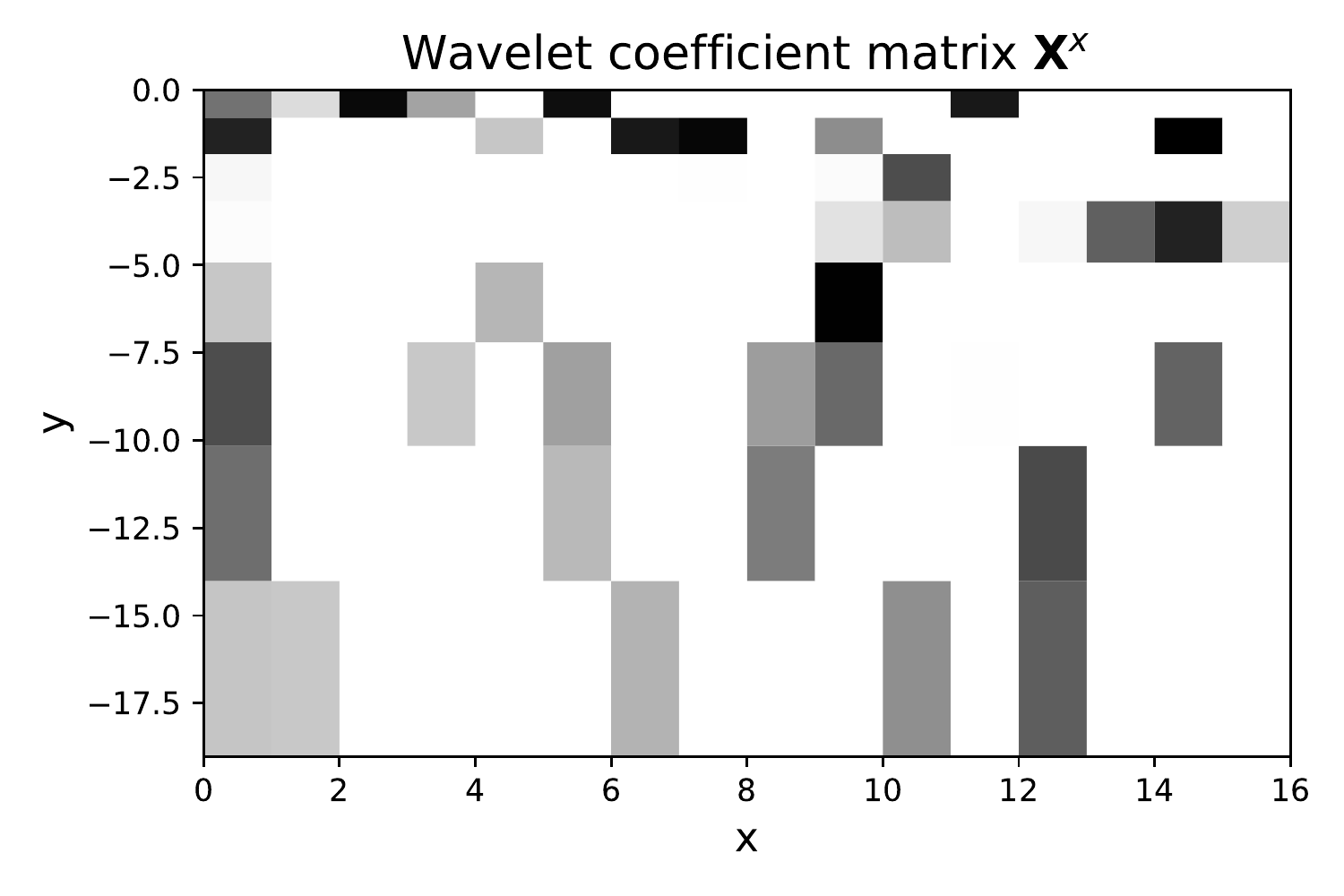}  \\
					E. \hspace{\widthFig\linewidth} F. \\
			\includegraphics[width=\widthFig\textwidth]{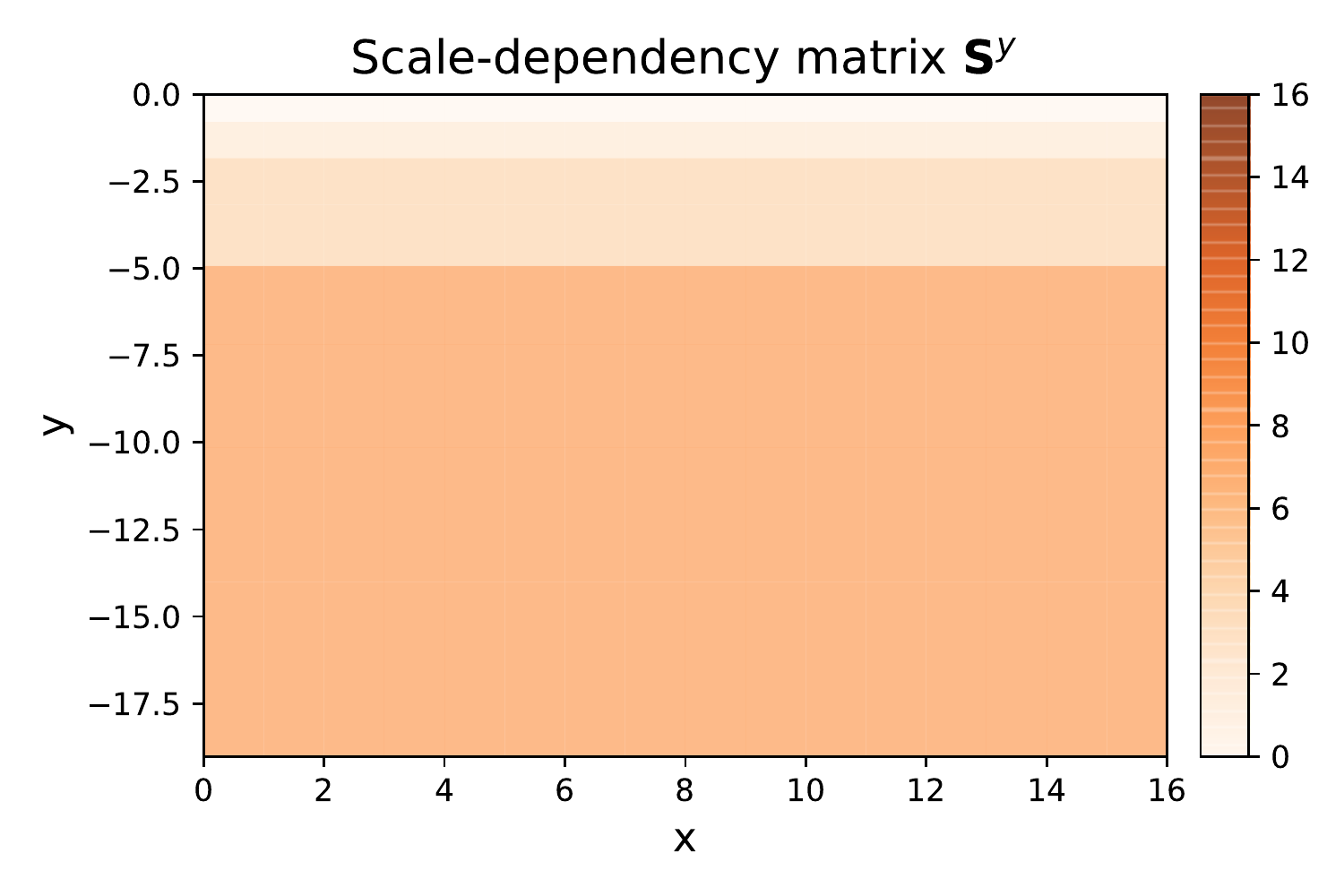}  \includegraphics[width=\widthFig\textwidth]{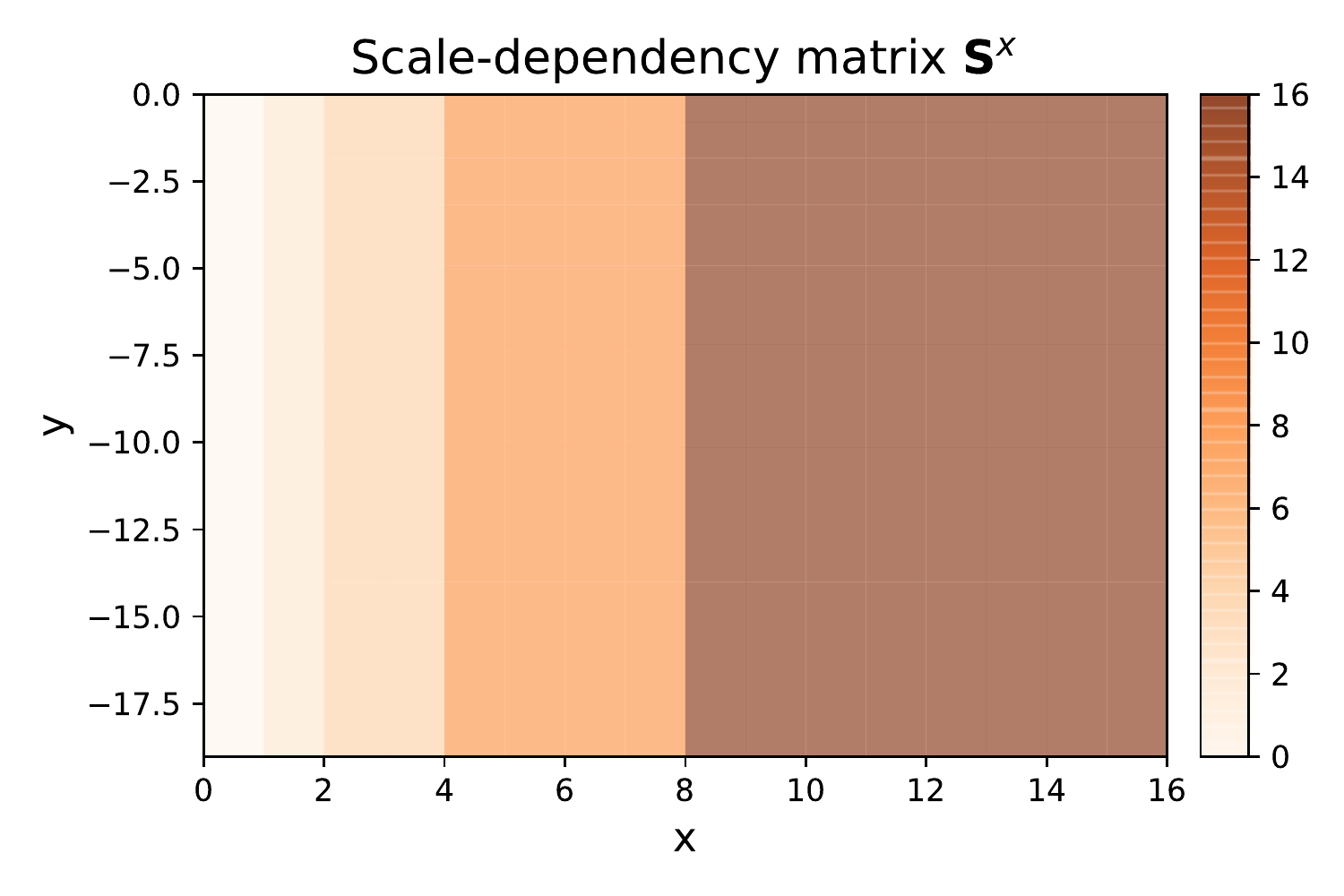}  \\
					G. \hspace{\widthFig\linewidth} H. \\
			\includegraphics[width=\widthFig\textwidth]{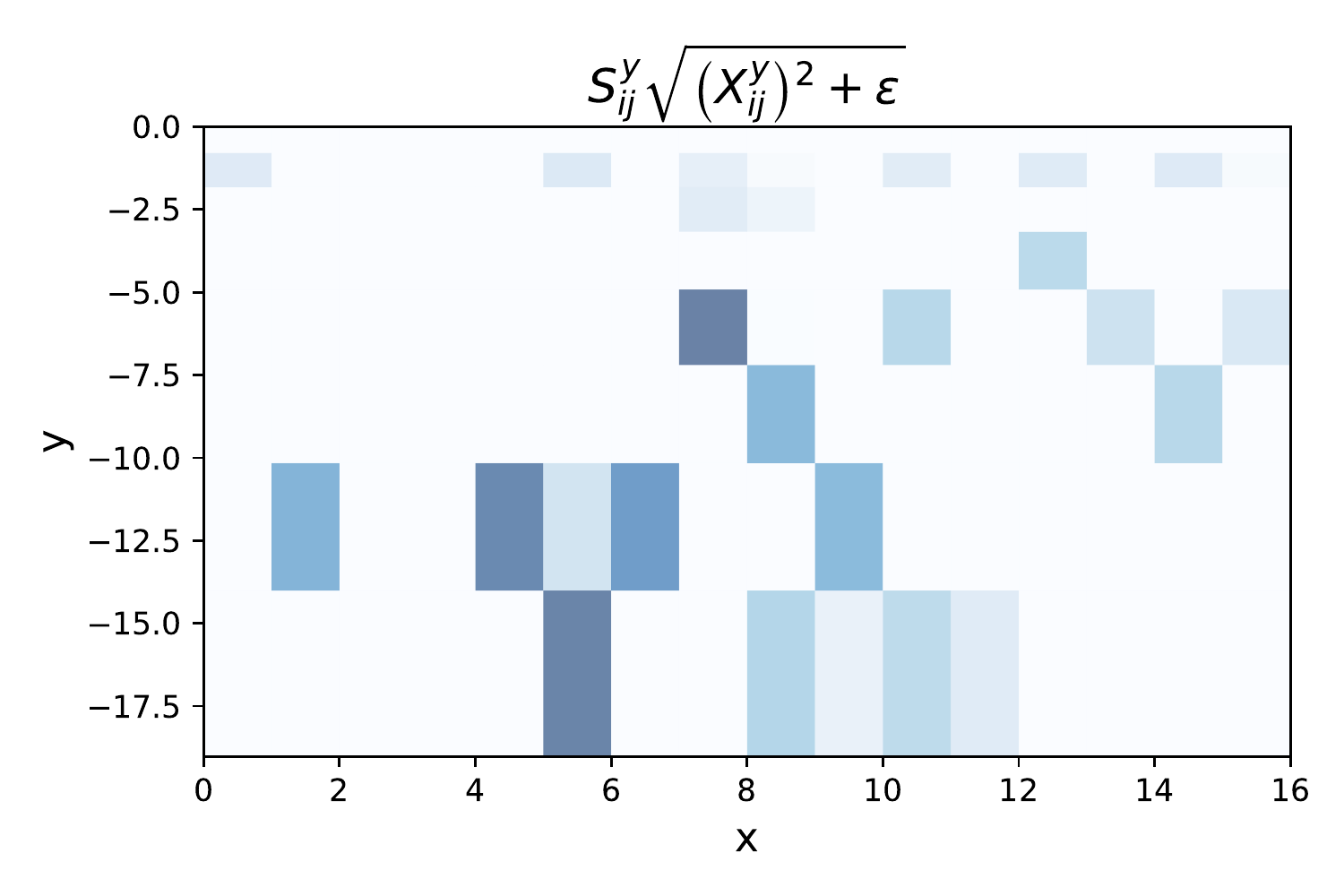}  \includegraphics[width=\widthFig\textwidth]{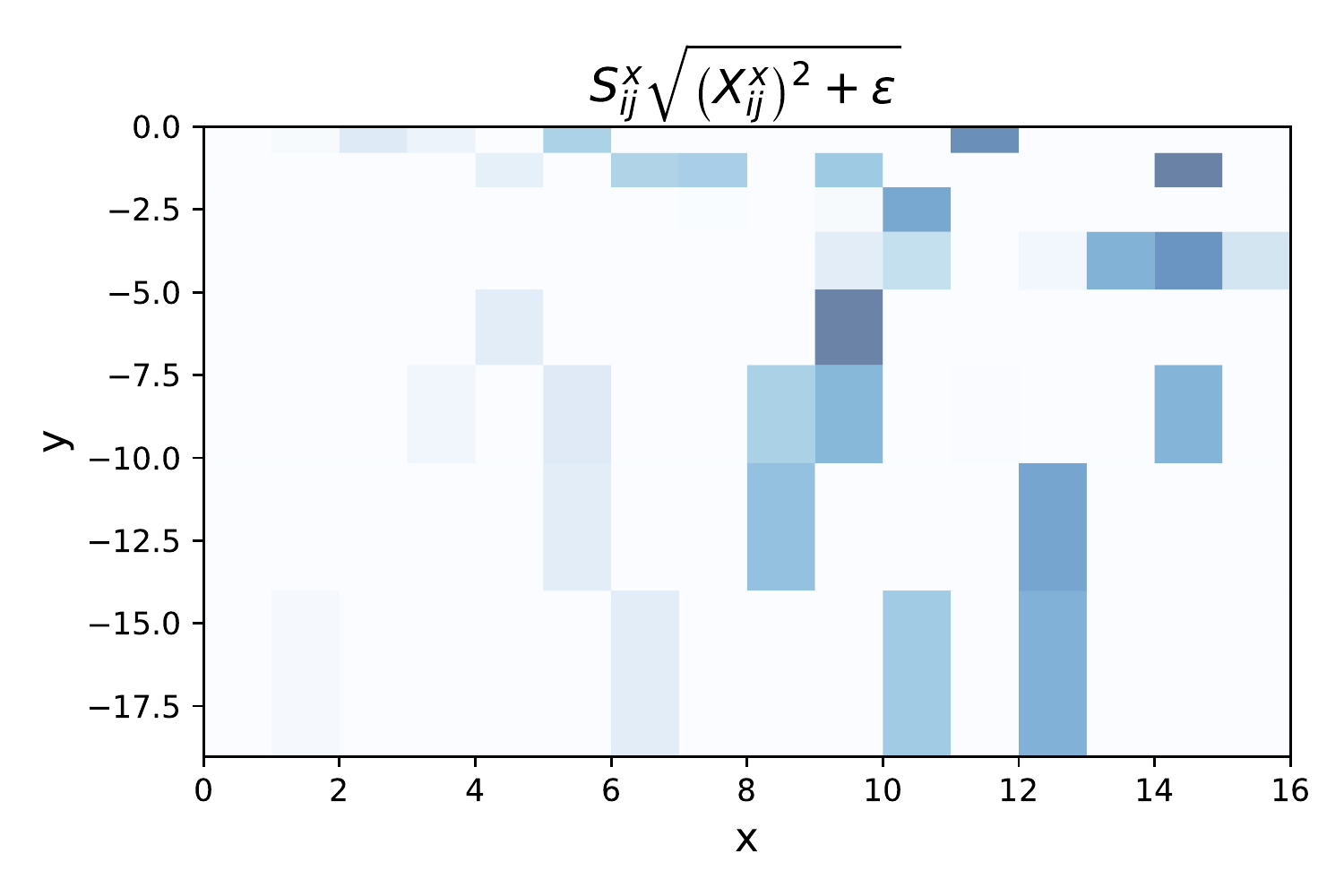}  \\
			
	\end{tabular}
\caption{Conceptual visualization of the consecutive steps towards the scale-dependent wavelet-based model misfit. The inversion model $\vb{M}$ with a discretization is sliced in vertical and horizontal snippets (A. and B.). On each snippet, the 1D discrete wavelet transform is applied. The wavelet coefficients in C. and D., together with the scale-dependency matrices E. and F. are used in the perturbed Ekblom measure in G. and H.}
\label{fig:conceptual}
\end{figure*}
\begin{figure*}
	\includegraphics[width=\linewidth]{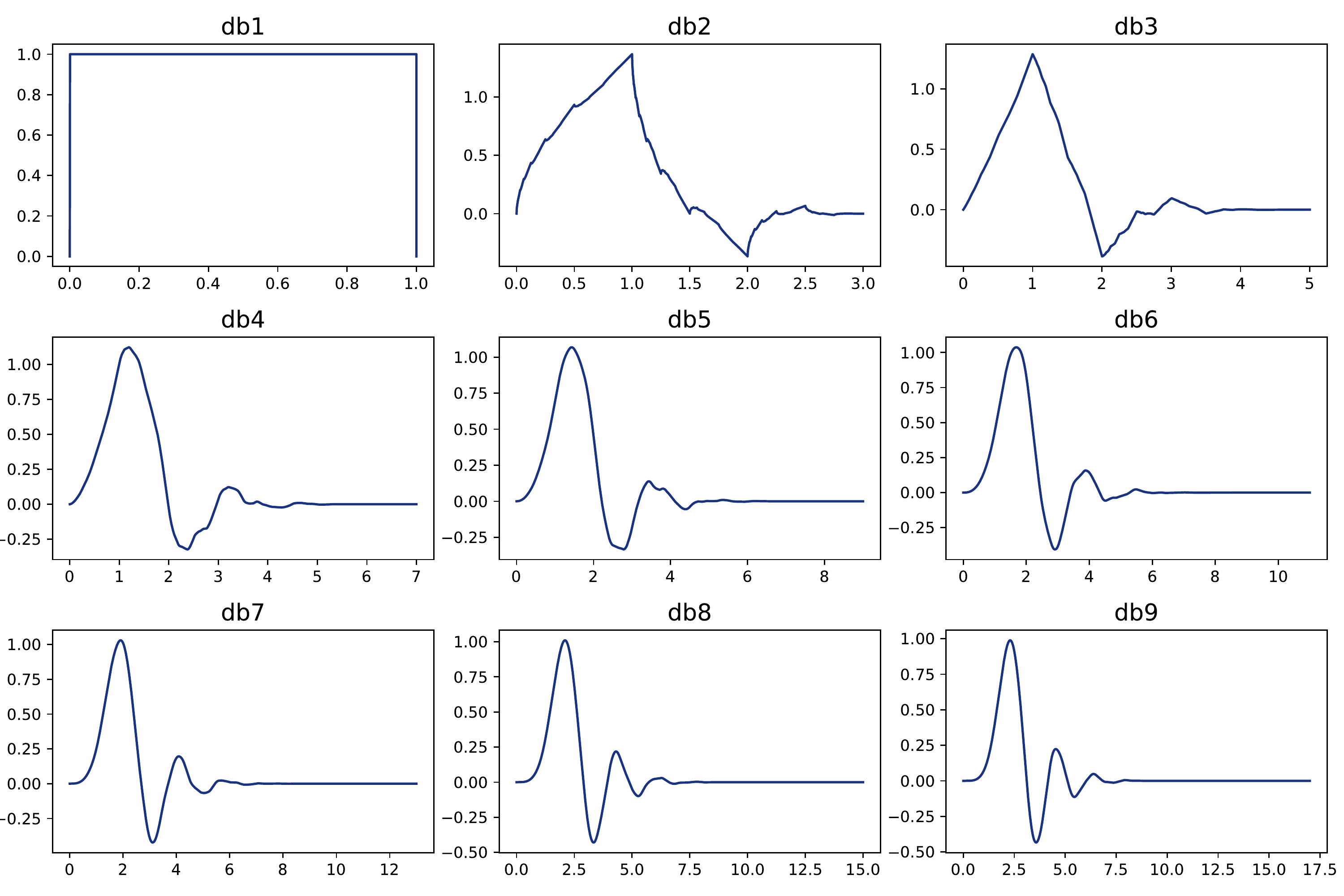}
	\caption{Daubechies mother wavelets $\varphi$. The characteristics (blocky, smooth, sharp) of the basis function determine the minimum-structure in the inversion model.}
	\label{fig:mother_wavelets}
\end{figure*}
\subsection{Scale-dependent wavelet-based complexity measure}
In this section, we extend the work of  \citet{deleersnyder2021inversion} into two dimensions. The 2D scale-dependent wavelet-based regularization term will be described in a step-by-step manner. An inversion model $\vb{M} \in \mathbb{R}^{n_m \times n_s}$ is a matrix, where $n_m$ is the number of layers and $n_s$ is the number of soundings, an example with an illustrative discretization is shown in Figures \ref{fig:conceptual}A and B. That inversion model $\vb{M}$ is sliced along both vertical and lateral orientation as 1D layer snippets $\vb{m}$, as demonstrated in Figure \ref{fig:conceptual}A-\ref{fig:conceptual}B. The model misfit $\phi_m$ splits into a sum of model misfits per orientation: 
\begin{equation}
		\phi_m(\vb{M}) =\phi_{m, \updownarrow}(\vb{M}) +  \alpha\phi_{m, \leftrightarrow}(\vb{M}),
\end{equation}
where $\alpha$, the relative regularization parameter, sets the relative importance of those terms.

\subsubsection{Measuring model complexity with the discrete wavelet transform}
\label{subsec:dwt}
\begin{figure*}
	\includegraphics[width=\linewidth]{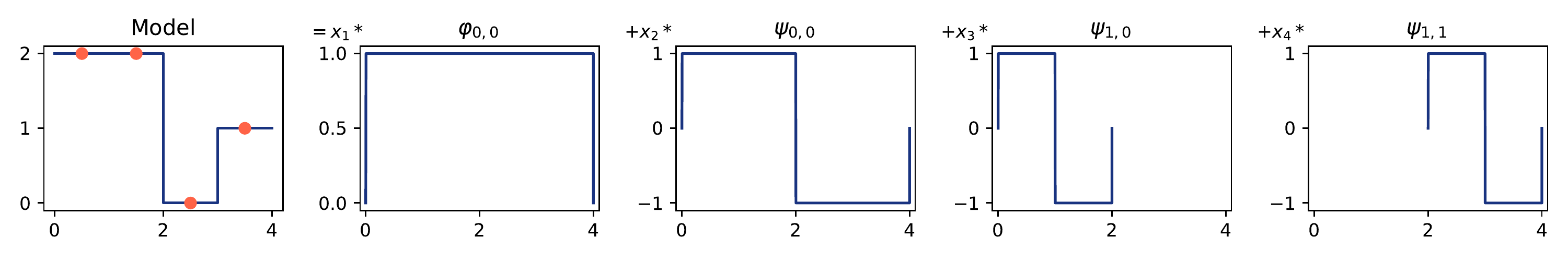}
	\caption{A simple example of a wavelet decomposition of a model with four model parameters into four wavelet basis functions. The model $\vb{m} = 1.25 \varphi_{0,0} + 0.75\psi_{0,0} + 0\psi_{0,0} + 0.5\psi_{1, 1}$ (the coefficients of the true wavelet transform are rescaled with $\sqrt{2}/2$ for each extra level, due to a scaling factor that was omitted in Eq. \eqref{eq: wavelets} and here for clarity).
}
	\label{fig:waveletdecomp}
\end{figure*}
Suppose that there exists a basis in which the true model parameters $\vb{M}$, known to possess minimum structure, are represented in a sparse fashion. Such a basis in combination with a sparsity promoting measure would yield an appropriate model misfit. This would also mean that a randomly generated model (or a heavily over-fitted model that one typically gets by setting the regularization parameter $\beta$ too low) would also be random in that basis. This would be heavily penalized by the sparsity promoting measure and therefore $\vb{M}$ would not be a minimum of $\phi$.\\

The discrete wavelet transform can be interpreted as a basis transformation $\vb{W}$ which transforms an inversion model snippet $\vb{m}$ from the model domain into the wavelet domain with coefficients $x_i$. The discrete
wavelet transform allows to represent the conductivity profile (or horizontal cross-section) in a sparse fashion, because it has both spatial and temporal resolution (the nomenclature temporal stems from the relation to the frequency domain, i.e. width of the signal, while spatial is related to the location of the wavelet basis function). It is intuitive to compare the wavelet transform with the Fourier transform. In Fourier analysis, we represent a signal/model in time/model domain and map it to the frequency domain. The original time signal can be viewed as a sum of Fourier coefficients and their basis functions $\{e^{ikt}\}_{k \in \mathbb{Z}}$. In wavelet theory, there are multiple available basis functions (see Figure \ref{fig:mother_wavelets}). The discrete wavelet transform is a sum of wavelet coefficients multiplied with their basis functions, which vary not in frequency but in width and location. 
A simple example is presented in Figure \ref{fig:waveletdecomp}, where a snippet of the inversion model with 4 model parameters has been decomposed in the wavelet domain. In this example, Daubechies 1 or Haar wavelets are used. The scaling function $\varphi_{0,0}$ (or mother function) is simply a block function over the whole domain of the model $\vb{m}$. The other wavelet functions are built from this mother function, that is a wavelet function $\psi_{0,0}$ and two wavelet functions with smaller compact support $\psi_{0, 1}, \psi_{1, 0}$. The relations between the wavelet function are as follows:
\begin{equation}
	\label{eq: wavelets}
	\psi_{n,k}(t) \sim	\psi(2^n t - k),
\end{equation}
where $\psi(t)$ is the wavelet basis function $\psi_{0,0}$.
The first parameter is related to the compact support width. The lower the $n$, the wider the compact support width. The $k$ parameter controls the translation. So each wavelet function gauges the model with a specific temporal and spatial resolution. One can always find wavelet-coefficients $x_i, $ where $i \in [1,2,3,4]$, such that the aforementioned model is exactly represented in terms of those four basis functions. It is evident that coefficient $x_3$ corresponding to basis function $\psi_{1, 0}$ should be zero for this specific example. This is the foundation of the sparsifying nature of the wavelet transform. A more detailed example is found in \citet{deleersnyder2021inversion} or in standard works about wavelet theory, such as \cite{mallat1999wavelet}.\\

The discrete wavelet transform is reliably computed via the Fast Wavelet Transform, as e.g., implemented in the PyWavelets package \citep{lee2006pywavelets}. It boils down to a matrix multiplication on each 1D snippet $\vb{m}$, which generates the wavelet representation $\vb{x}$. Repeating this on each snippet generates a matrix $\vb{X}$ with wavelet representations for each orientation:
\begin{equation}
	\vb{X}^x= \vb{W}^x\vb{M}, \qquad  \vb{X}^y = \vb{MW}^{y\text{T}},
\end{equation}
where each orientation has its own basis transformation matrix $\vb{W}^{x/y }$. A conceptual illustration of such sparse wavelet coefficient matrices $\vb{X}$ are shown in Figure \ref{fig:conceptual}C-\ref{fig:conceptual}D.

\subsubsection{Choice of the wavelet}
There exist many different types of wavelet basis functions. Wavelet basis functions with similar properties are grouped in wavelet families. Those wavelets within a family are often developed to possess specific characteristics in shape but are more often the result of possessing some other mathematical properties (symmetry, orthogonality, minimal compact support...). The shape of the wavelet is often a corollary. The Haar wavelet (already used in Figure \ref{fig:waveletdecomp}) is likewise a wavelet from the Daubechies family. This family is by far the best-known and will be the only family used here. The Haar wavelet is also known as the Daubechies one (db1) wavelet. The `one' refers to the number of `vanishing moments'. This decisive property of a wavelet is defined as follows:\\
A wavelet has $p$ vanishing moments when 
\begin{equation}
	\int t^k \psi(t) dt = 0 \qquad \text{for} \quad k = 0, 1, ... , p-1. 
	\label{eq:vanishing_moments}
\end{equation}
The number of vanishing moments is related with the compact support of the wavelet: an orthonormal wavelet with $p$ vanishing moments has at least a support of size $2p - 1$ \citep{daubechies1988orthonormal}. The Daubechies family is defined (by imposing mathematical properties rather than on appearance, cf. supra) as the set of wavelets with minimal compact support, given a number of vanishing moments $p$. All the mother wavelet basis functions that are being used in this work are listed in Figure \ref{fig:mother_wavelets}.\\

The Daubechies family is known for having good approximating abilities, this means that it can represent many piecewise smooth models in a sparse fashion (see Section \ref{sec:scale-dep}). The number of vanishing moments also has an effect on the sparsity of an inversion model. Equation \eqref{eq:vanishing_moments} implies that a wavelet with $p$ vanishing moments is orthogonal to polynomials of degree $p-1$. Hence, the db2 wavelet is orthogonal to linear functions  and this guarantees that wavelet-coefficients will be zero for linear pieces in an inversion model $\vb{m}$. The greater the number of vanishing moments, the more complex structures can be represented in a sparse fashion.
\subsubsection{Scale-dependent complexity measure}
\label{sec:scale-dep}

We introduce specific scale-dependency matrices $\vb{S^x}, \vb{S^y}$ which depend on the structure of the wavelet coefficient matrices $\vb{X^x}, \vb{X^y}$. The entries of $\vb{S}$ are zero if it corresponds to coefficients of the mother wavelet or scaling function in $\vb{X}$ (those coefficients determine the `total energy' of the model, in other terms if the scaling coefficients are zero, the integral over the model should be zero). This is because those coefficients must never be zero for a realistic inversion model and should therefore not be minimized. Note that in the conceptual Figure \ref{fig:conceptual}E and \ref{fig:conceptual}F, all scaling function coefficients are \textit{not} zero. The other entries of the scale-dependency matrices $\vb{S}$ are $2^n$, where $n$ is the dilation parameter of the corresponding entry in $\vb{X}$. The scale-dependency matrices $\vb{S^x}, \vb{S^y}$ of our conceptual example are shown in Figures \ref{fig:conceptual}G and \ref{fig:conceptual}H.\\

In \citet{deleersnyder2021inversion}, we compare our scale-dependent regularization scheme with a different wavelet-based regularization scheme without this scale-dependency feature. A main advantage is that the scale-dependency of our scheme allows the use of wavelets with few vanishing moments and is, therefore, an improvement with respect to existing wavelet-based regularization schemes which are mostly successful with wavelets with larger vanishing moments (and thus smooth models). The idea behind the scale-dependency can be understood more intuitively with the building block metaphor, sketched in Appendix \ref{ap:buildingblock}.

\subsubsection{The perturbed $\ell_1$-Ekblom measure}
The presented minimum-structure regularization scheme is based on sparsity. The perfect sparsity measure would be the $\ell_0$-``norm", which is in fact a quasi-norm. It counts the number of non-zero entries in a vector. However, the $\ell_0$-``norm" is impractical in optimization. \citet{donoho2006most} has shown that the $\ell_1$-norm is a good approximation for the $\ell_0$-``norm" and thus a reliable sparsity promoting measure. \\

The $\ell_1$-norm is not differentiable at zero, therefore the perturbed $\ell_1$-norm measure of Ekblom \citep{ekblom1987l1} is used. The measure
\begin{equation}
	\mu_{\mathrm{Ekblom}}(x) = \sqrt{ x^2 + \epsilon}
\end{equation}
reduces to the $\ell_1$-norm for vanishing $\epsilon$ and is also convex. In our scheme, wavelet coefficients $x_i$ smaller than $10^{-4}$ have little effect on the actual conductivity profile in the model domain. Further tests on the sensitivity of $\epsilon$ reveal that there is only a negligible effect of the value of $\epsilon$ on the inversion result. In this paper, $\epsilon$ is always set equal to $10^{-6}$.\\

In Figures \ref{fig:conceptual}G-\ref{fig:conceptual}H, the model misfit, before summation, is shown. The darker, the larger the cost of the entry. Note that due to the scale dependency, the wavelet coefficients corresponding to coarser wavelet basis functions are lighter and that coefficients from the scaling functions do not contribute at all.

\subsubsection{Calibration: Relative weighting parameter $\alpha$}
\label{sec:calibration}
In summary, the scale-dependent wavelet-based model misfit term in two dimensions is
\begin{align}
		\phi_m(\vb{M}) &= \phi_{m, \updownarrow}(\vb{M}) +  \alpha\phi_{m, \leftrightarrow}(\vb{M})\\
								& = \sum_{i,j} S^x_{ij}\sqrt{ \left(X^x_{ij}\right)^2 + \epsilon} + \alpha \sum_{i,j}S^y_{ij}\sqrt{ \left(X^y_{ij}\right)^2 + \epsilon},
\end{align}
where $X^x_{ij} = W^xM$,  $X^y_{ij} = W^yM^\text{T}$, and $\alpha$ balances the relative weight of both individual model misfit terms. With an extra calibration step, the parameter $\alpha$ can be estimated. This can be the final parameter $\tilde{\alpha}$, or it can be chosen as the initial $\tilde{\alpha}^0$ in a more comprehensive optimization strategy (in our examples, we demonstrate this with a sweep centered around $\tilde{\alpha}^0$). If we deviate from $\tilde{\alpha}$, we label this with $\gamma$, such that 
\begin{equation}
	\alpha = \gamma \tilde{\alpha}.
\end{equation}

For AEM data we know that early time data is sensitive to the upper layers of the inversion model and that late time data probes deeper layers. Moreover, one expects that the response increases if the electrical conductivity increases, so the raw TDEM data in itself gives an insight into the amount of structure/complexity to expect in inversion models. As an extreme case, for horizontally stratified inversion models you have no lateral variation in the (noise-free) data, while a sharp, lateral transition in the data will correspond to a sharp, lateral transition in the inversion model. We will use this characteristic behaviour to estimate the size of the horizontal model misfit (see below).\\

We propose the parameter $\tilde{\alpha}$ as the ratio of the estimation of the model misfit in the vertical orientation $\tilde{\phi}_{m, \updownarrow}$ by the model misfit of the horizontal orientation $\tilde{\phi}_{m, \leftrightarrow}$:
\begin{equation}
	\tilde{\alpha} = \frac{\tilde{\phi}_{m, \updownarrow}}{\tilde{\phi}_{m, \leftrightarrow}}.
\end{equation}

$\tilde{\phi}_{m, \updownarrow}$ is estimated by randomly picking a sounding and to perform a 1D inversion with the appropriate wavelet basis function (up to $\phi_d = 1$). The model complexity is then extrapolated on each sounding and thus 
\begin{equation}
	\tilde{\phi}_{m, \updownarrow} = n_s \cdot \phi_{m, \text{1D}}.
\end{equation}
Clearly, this estimation can be improved by averaging over 1D model complexities of multiple soundings. In this work, only one sounding is used for the estimation of $\tilde{\phi}_{m, \updownarrow}$.\\

The lateral variations in the TDEM data are transformed to variations in model parameters-like values (ranging between 0 and 1). The horizontal model misfit $\phi_{m, \leftrightarrow}$ on the transformed TDEM data then captures the expected complexity. The first step in the transformation is a multiplication with a normalization factor. For the data $d^i$ at time channel $i$, the normalized data $\tilde{d}^i $ is computed as
\begin{equation}
	\delta^i = \frac{d^i}{ \max(d^i) -  \min(d^i)} 
\end{equation}
\begin{equation}
	\tilde{d}^i =\delta^i - \min(\delta^i) + 10^{-9},
\end{equation}
where the $\min(\delta^i)$ translates the negative data to positive values (the value for electrical conductivities are always positive) and $10^{-9}$ ensures strict positiveness for the log-transform (see later). Then the lateral model misfit is computed on the data (each time channel corresponds to a horizontal snippet). This is then corrected for the total number of snippets in the inversion model, i.e., divided by the number of time channels and multiplied with the number of layers in the discretization of the sounding $n_m$.\\

We highlight that the motivation behind the calibration step is twofold: the first and most important reason is that $\phi_{m, \updownarrow}(\vb{M})$ and $\phi_{m, \leftrightarrow}(\vb{M})$ can differ greatly, due to multiple reasons. One cause can be that the number of soundings is much larger than the number of vertical model parameters. Another significant cause is the choice of the wavelet basis. A snippet $\vb{m}$ can generate a much larger model misfit in one wavelet basis than another, due to the different number of levels in the wavelet transform (a wavelet basis function with a larger number of vanishing moments has less levels in the wavelet transform) and increased by the scale-dependency from Section \ref{sec:scale-dep}, which assigns larger weights and eventually results in larger model misfits. Since we want to potentially generate all different wavelet-wavelet combinations, without manually picking an $\alpha$, this automation step is crucial. The second reason for this calibration is that, thanks to the use of complexity in the AEM data, the calibration can also estimate an (initial) value for the relative regularization parameter $\alpha$, `knowing' about the expected lateral complexity.
\subsection{The forward model}
\label{sec:forward}
The forward model $\mathcal{F}$ describes the subsurface response to a magnetic dipole, given the parameter distribution of the subsurface and the set-up of the measurement instrument. There exist two main types of forward models: (1) (semi-)analytical forward models that solve the (continuous) Maxwell equations and (2) simulations based on discretization of the physics. To mimic the full 3D subsurface response of the potentially non-1D subsurface, one resorts to 3D simulations of the physics. However, for geophysical inversion this is computationally intractable, primarily because of the computation of the Jacobian. \\

In this work, we resort to the semi-analytical solution by \citet{hunziker2015electromagnetic}. It derives the magnetic field response for a 1D layered earth in the wavenumber-frequency domain in a more general fashion than \citet{wait1951magnetic}. The transformation into space and time domain is via the Hankel and Fourier transform, respectively. An open-source Python implementation by \citet{werthmueller2017open} implements those equations in a fast and reliable fashion, allowing to select between the most common Fourier and Hankel transform methods. In this paper, we have employed the digital linear filters by \citet{key20091d} (Key 81 pt cosine sine filter and Key 201 pt Hankel filter). They are a more recent version of the digital filters behind the fast Hankel transform by \citet{anderson1979numerical} and have proven to be very fast and precise for frequencies in the range of CSEM data \citep{werthmueller2017open}. 

\section{Results}
\label{sec:results}
In this section, we demonstrate the flexibility of our regularization scheme, that is the potential to recover blocky, smooth, and intermediate inversion models and the combination of blocky shapes in one direction and smoother transitions in the other. By a comparsion with the true model, we discuss the results both qualitatively and quantitatively. In our analysis, an inversion with the LCI method \citep{auken2004layered} is added, which is a common AEM inversion technique, showing that our flexible method can generate comparable results, provided a specific choice of wavelet-wavelet basis functions. \\

We have selected a synthetic model in a similar context to our field data case of Section \ref{sec:fielddata} such that we can use the specific conclusion to our case study. The geophysical survey has an identical set-up as the real field data case, that is time-domain AEM data obtained with a SkyTEM's 304M system, which has a dual moment central-loop configuration. The Low Moment (LM) has a higher resolution in the shallow subsurface, because the lower moment can be turned off comparably faster and therefore the early time gates can be interpreted. The High Moment (HM) has a larger resolution at larger depths, due to its larger peak moment. Both the LM and HM have a non-square waveform (see \citet{vlaanderentopsoil} and the supplementary material). The transmitter loop is an octagon of 340.8 m$^2$, but this is approximated by a loop in the forward model. The receiver coil measures the $z$-component of the magnetic field and has an effective receiver area of 105 m$^2$. It is located 2 m above and 13.20 m behind the centre of the transmitter loop. The height of the loop w.r.t. the surface varies with an average of 40.8 m (15.2 m std) and is measured with an altimeter. This variation is taken into account, while the tilt has been neglected. In the synthetic data case, the altitude is fixed to 40 m.



\subsection{Synthetic data case}
\label{sec:synthetic}
\begin{figure}
		\centering 
	\includegraphics[width=0.5\textwidth]{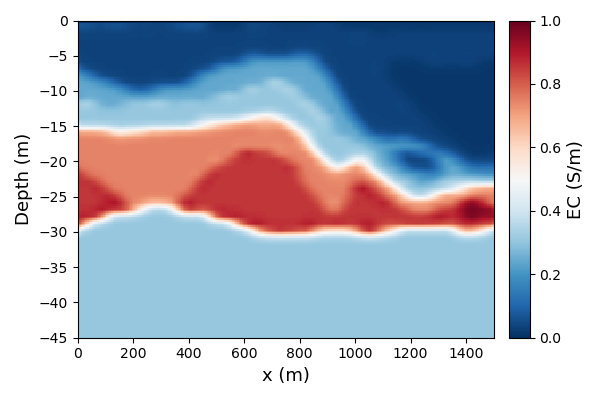}
	\caption{Synthetic true model (adapted from \citet{lebbe1986salt}).} 
	\label{fig:true_model}
\end{figure}

\begin{figure}
	\centering 
	\includegraphics[width=0.5\textwidth]{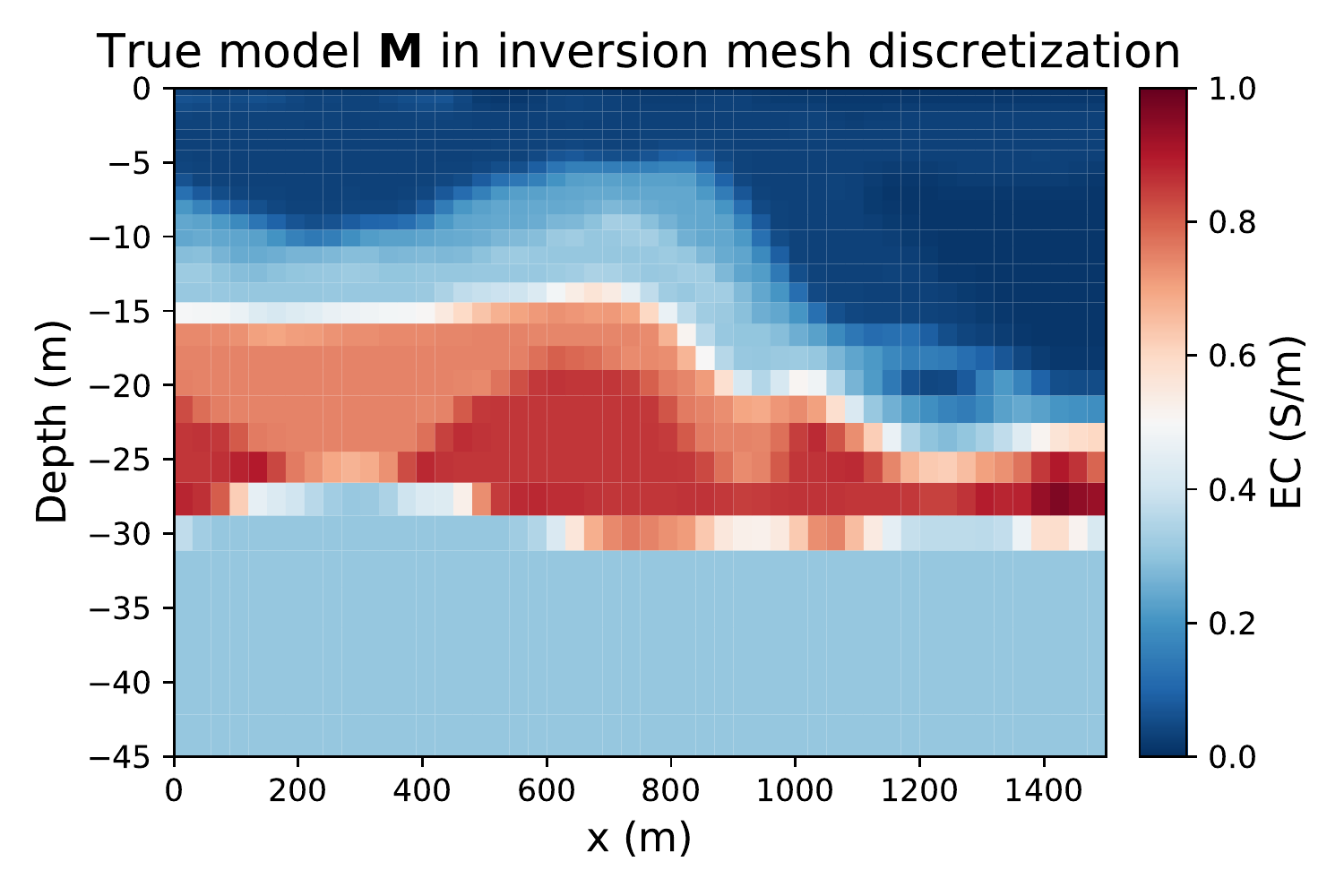}
	\caption{Synthetic true model (adapted from \citet{lebbe1986salt}) with the discretization of the inversion model.} 
	\label{fig:true_model_inversion_discretization}
\end{figure}

The synthetic model is shown in Figure \ref{fig:true_model} and is taken from \citet{lebbe1986salt}. It is a profile from the Belgian-French border which is covered by mudflats and where the upper part is freshwater that rests on a brackish bottom layer. The brackish water rises near two drainage ditches due to an upward flow (located at a distance $X$ of 0 and 600 m). The first layer ends at approximately 30 m and has a mostly sandy lithology, while at 30 m depth, the Kortrijk Clay formation is reached on which the saltwater rests. The synthetic model has a large variation in model parameter values, ranging from 5 mS/m to 1000 mS/m.\\

The synthetic data (see supplementary data) is generated with SimPEG's finite volume method forward modelling solver \citep{cockett2015simpeg, heagy2017framework} and the moving footprint approach \citep{cox20103d} which allows for parallelization, such that the data contains a multidimensional component to make the synthetic data case more realistic. Note that this forward model is different to the (much faster) 1D forward model used in the inverse problem. $3\%$ multiplicative noise is added to the synthetic data. There are 45 vertical model parameters, with equidistant spacing in $\log_{10}$-space (note that not all the model parameters are shown).\\

The discrepancy between 1D and 2.5D modelling turns out to be limited and accounts for a data misfit $\phi_d$ of 0.6 in the absence of the $3\%$ multiplicative noise. The average relative error is $-0.4\%$ and the standard deviation is $1.6\%$. When adding the $3\%$ multiplicative noise, the data misfit $\phi_d$ rises to 0.97, which is close to the root mean-squared error $\epsilon_{\text{RMS}}$ equal to one, as expected. When the true model is projected on the inversion mesh (that is the discretization used in the inversion), such as shown in Figure \ref{fig:true_model_inversion_discretization}, the data misfit $\phi_d$ rises to 1.37. \\
\begin{figure*}
	\includegraphics[width=0.45\linewidth]{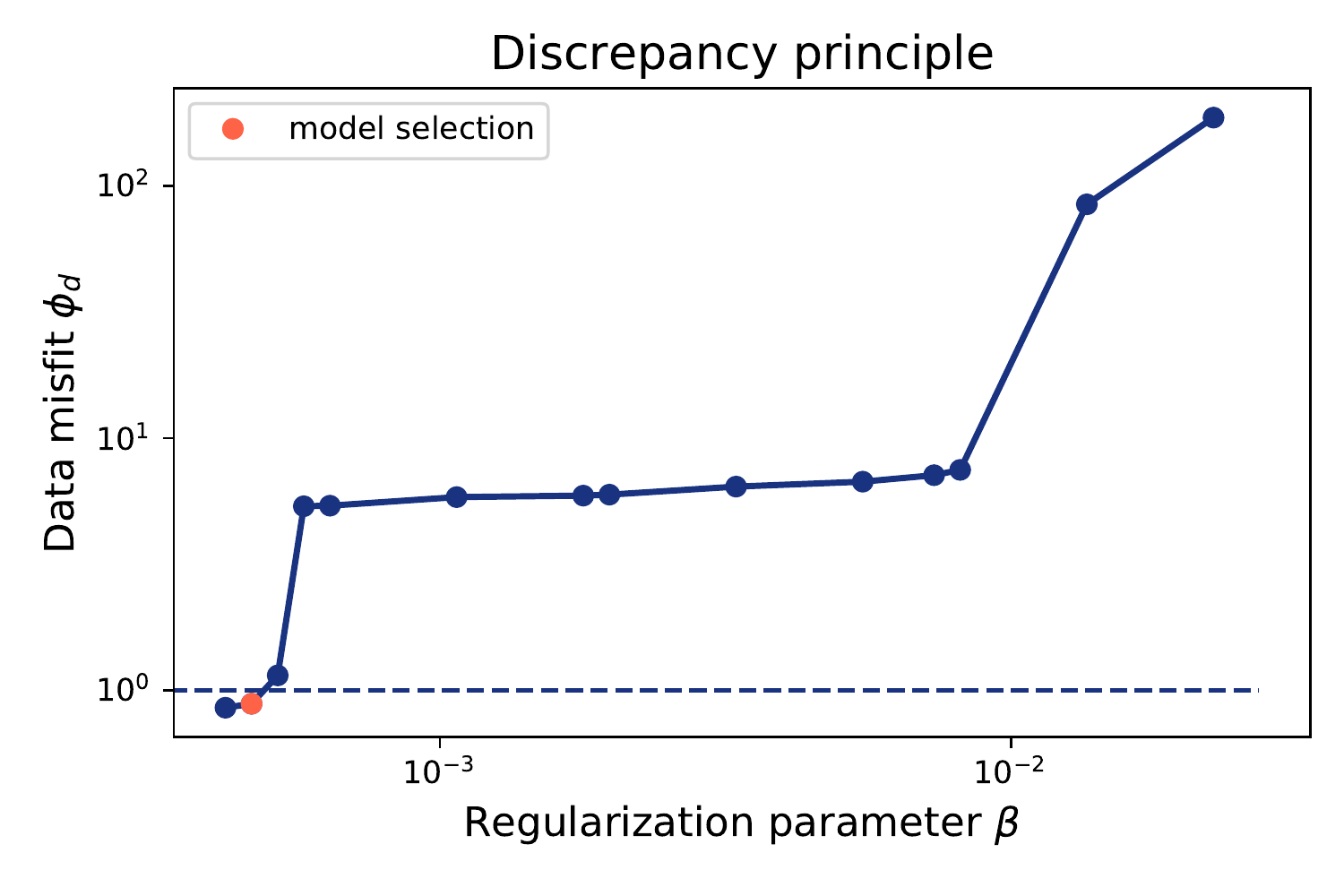}
	\hspace{2.5em}
	\includegraphics[width=0.45\linewidth]{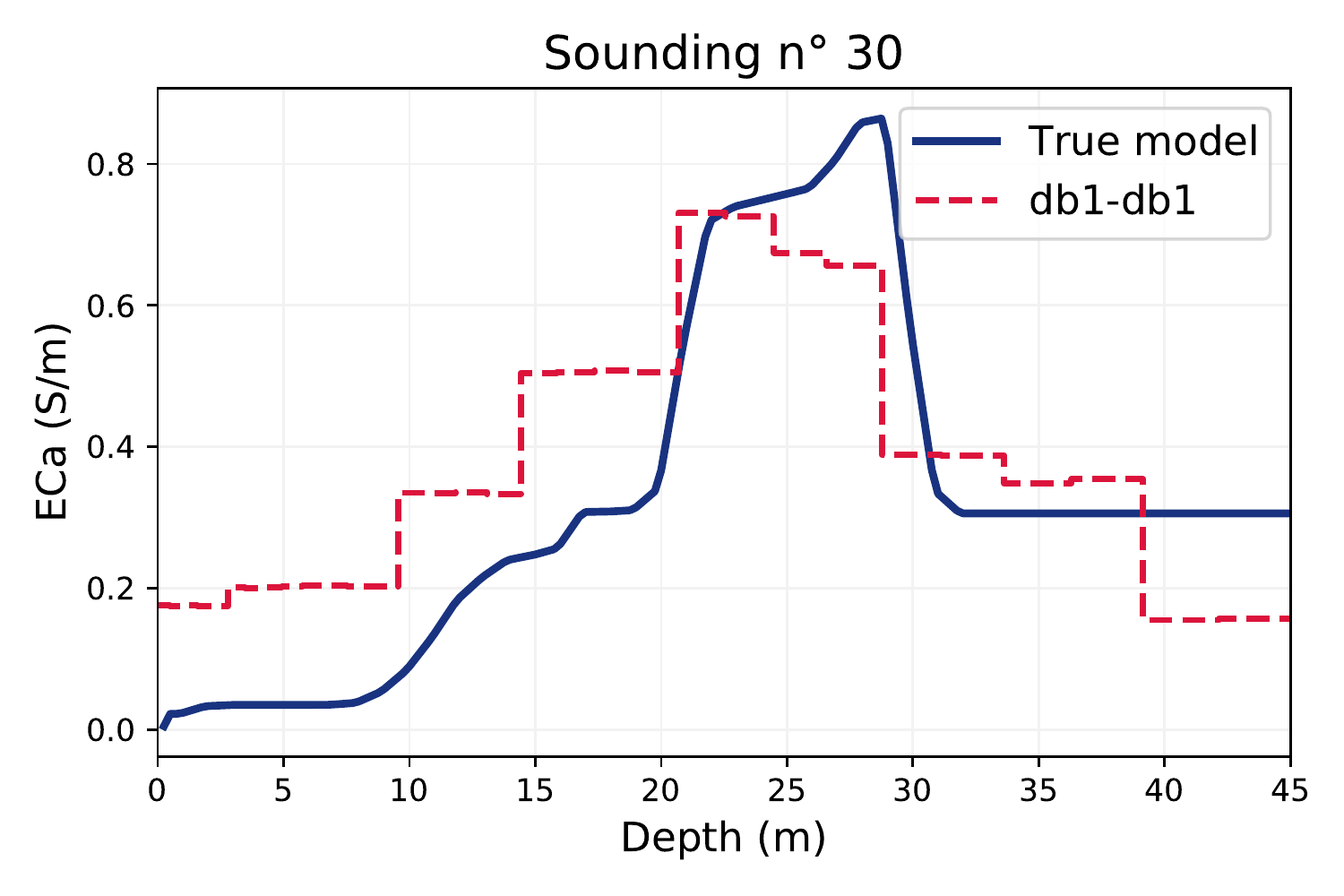}
	\includegraphics[width=0.45\linewidth]{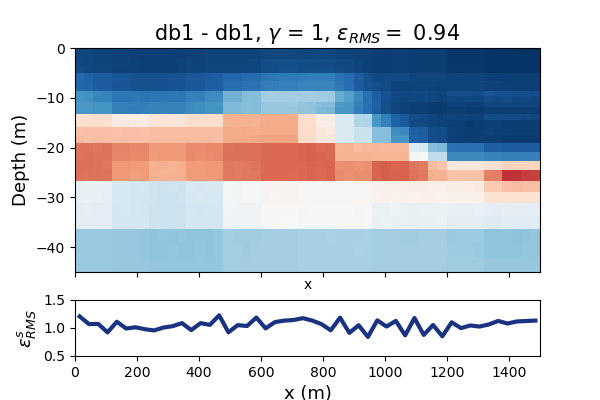}
	\hspace{-1.5em}\includegraphics[width=0.06\linewidth]{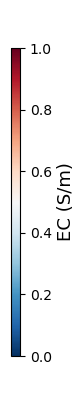}
	\includegraphics[width=0.45\linewidth]{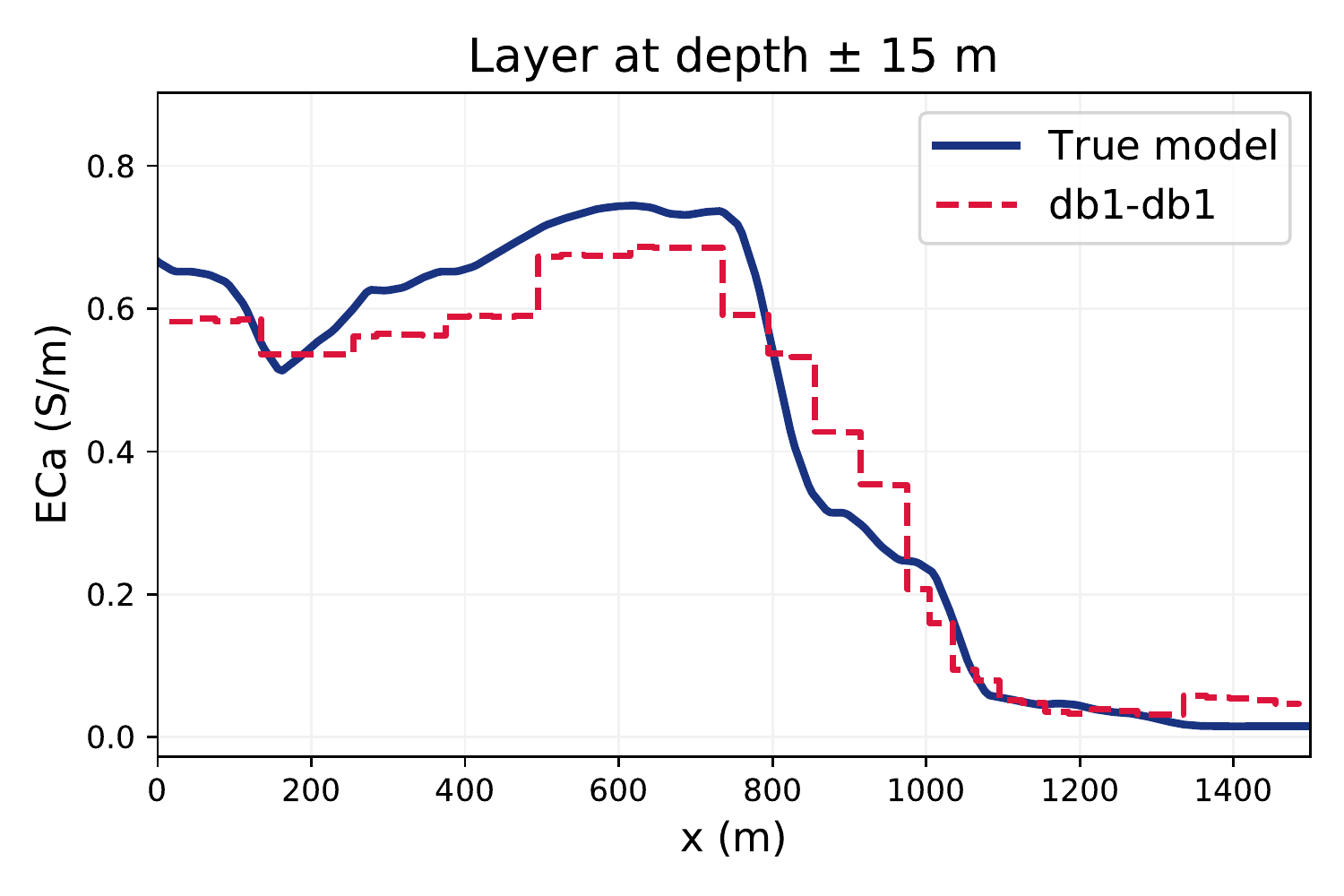}
	\caption{The inversion model. The discrepancy principle was used to determine the regularization parameter $\beta$. The calibration procedure estimated the value for the relative regularization parameter $\tilde{\alpha} = 0.5$. With $\gamma$ = 1, $\alpha = \gamma\tilde{\alpha} = 0.5$.} 
	\label{fig:V_11}
\end{figure*}
\begin{figure*}
	\begin{tabular}{lll}
		A. & B. & C.\\
		\includegraphics[width=0.33\linewidth]{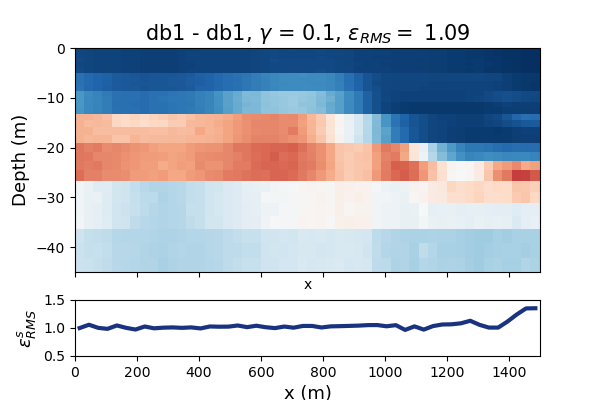}&\includegraphics[width=0.33\linewidth]{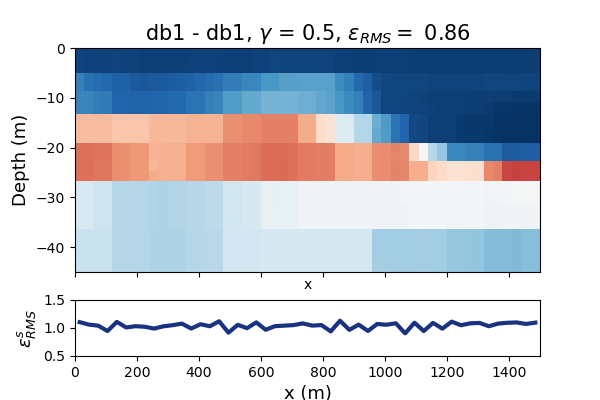} & \includegraphics[width=0.33\linewidth]{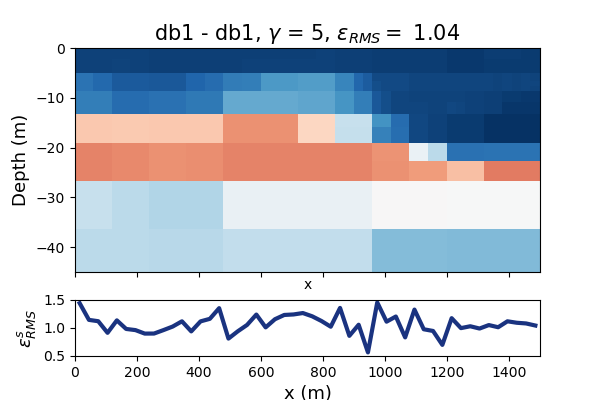}  \\ 
		D. & E. & F.\\
		\includegraphics[width=0.33\linewidth]{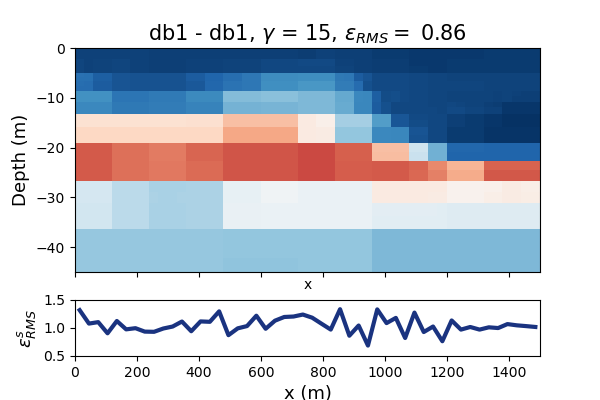}&\includegraphics[width=0.33\linewidth]{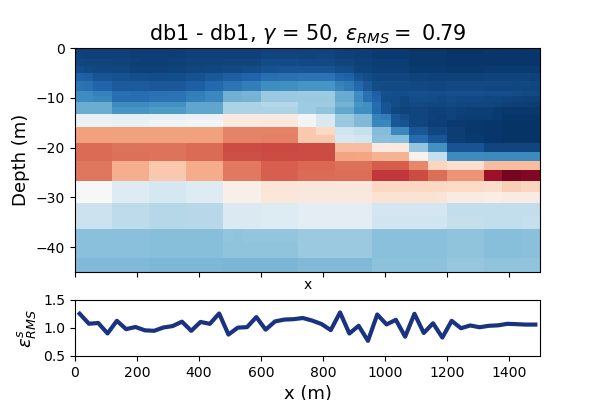} & \includegraphics[width=0.33\linewidth]{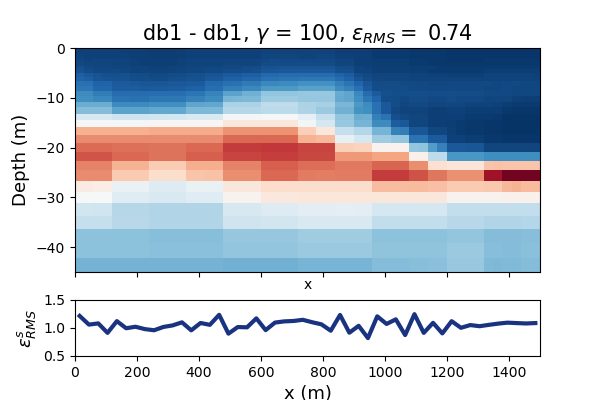}   \\ 
		
	\end{tabular}
	\caption{Inversion models for the db1-db1 wavelet-wavelet combination. A sweep over $\gamma$ was used to further optimize the estimate $\tilde{\alpha}$ via $\alpha = \gamma\tilde{\alpha}$. The noise weighted root mean-squared per sounding $\epsilon_{\text{RMS}}^s$ is shown below each inversion model.}
	\label{fig:sweep}
\end{figure*}
\begin{figure*}
	\begin{tabular}{lll}
		
		A.  & B.& C.\\
		\includegraphics[width=0.33\linewidth]{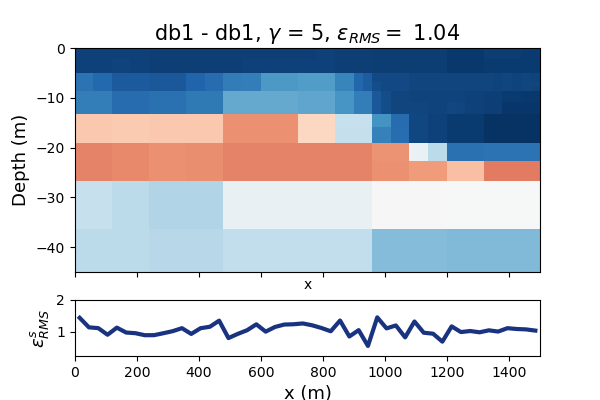}&\includegraphics[width=0.33\linewidth]{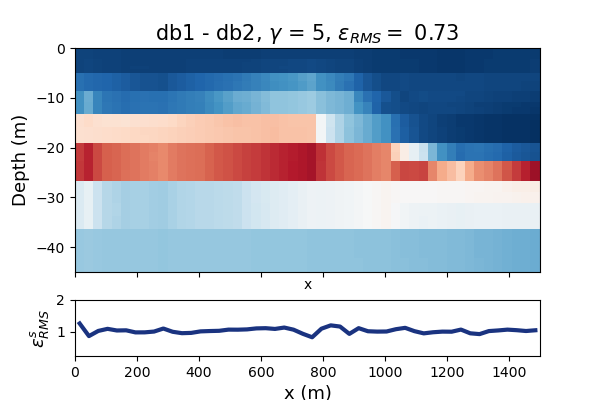} & \includegraphics[width=0.33\linewidth]{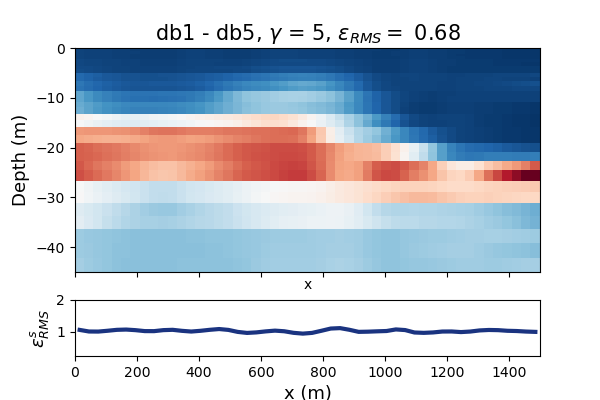}  \\ 
		
		D. & E. & F. \\
		\includegraphics[width=0.33\linewidth]{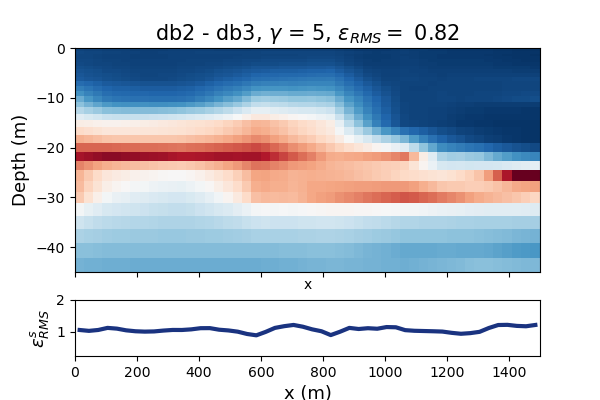}&\includegraphics[width=0.33\linewidth]{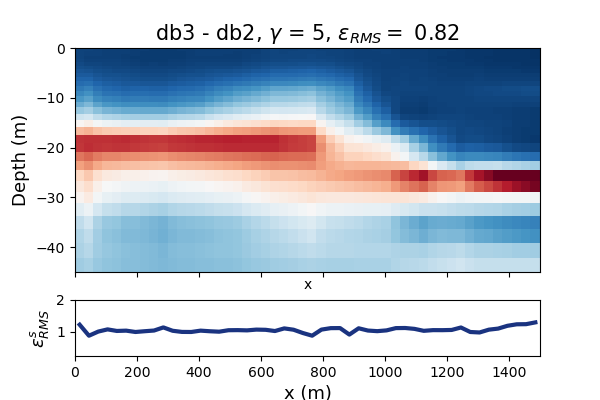} & \includegraphics[width=0.33\linewidth]{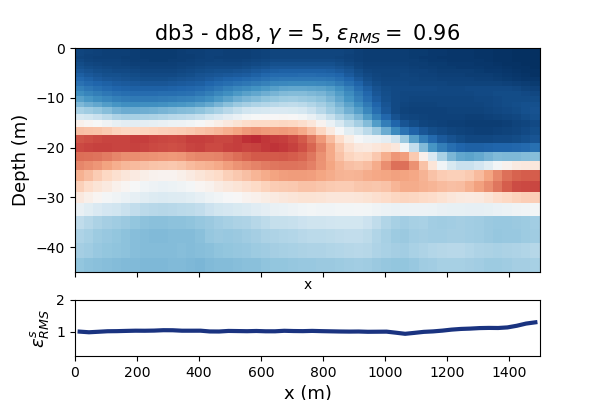}   \\ 
		
		G. & H. & I. \\
		\includegraphics[width=0.33\linewidth]{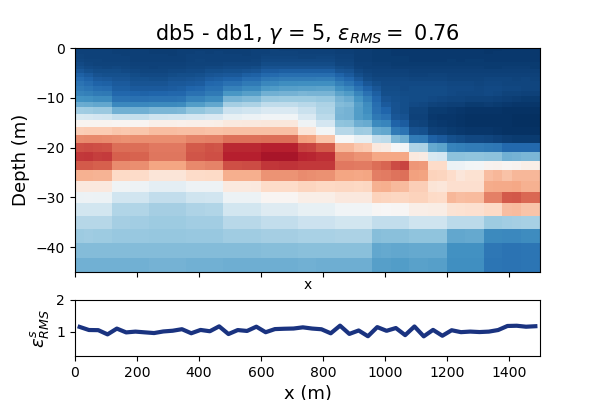}&\includegraphics[width=0.33\linewidth]{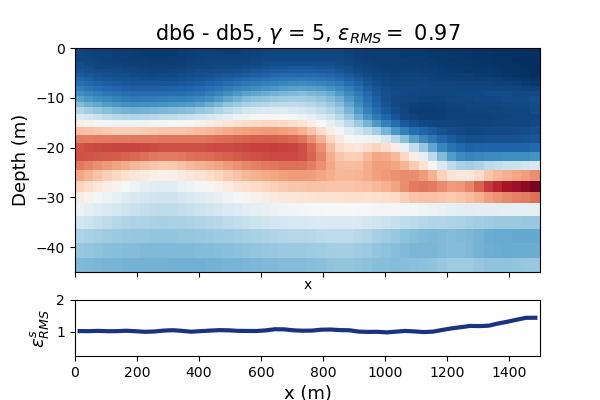} & \includegraphics[width=0.33\linewidth]{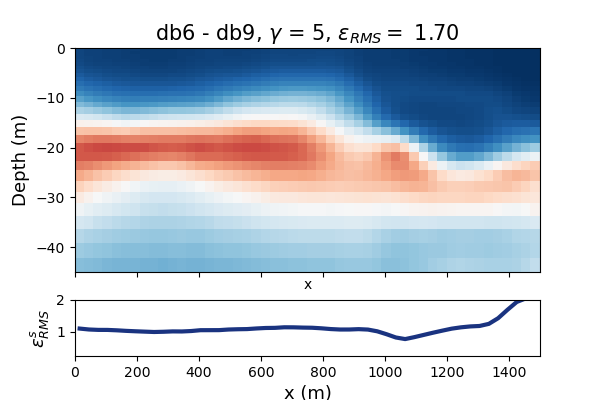}   \\ 
		
	\end{tabular}
	\caption{Selected models form the ensemble of inversion models for the synthetic data generated with the the true model in Figure \ref{fig:true_model}. db$N$-db$M$ wavelet combination refers to the db$N$ wavelet basis function used for the minimum structure along the vertical orientation, while db$M$ imposes the minimum structure on the horizontal orientation. Each inversion model uncovers different features from the synthetic data (and not necessarily of the true model).} 
	\label{fig:VI_2D}
\end{figure*}

The calibration-step estimates the relative regularization parameter $\tilde{\alpha}$ which are tabulated in Table \ref{tab}. Not all wavelet-wavelet combinations are presented, we focus on inversion models with significant distinct features.\\

After the calibration step, one inversion model is generated with $\gamma = 1 $ (or $\alpha = \tilde{\alpha}$) to assess whether the relative importance of the horizontal and vertical model misfit is geologically realistic.
In this example, the inspection is conducted with the db1 wavelet in both orientations. The result is shown in Figure \ref{fig:V_11}. In the discrepancy principle, we observe a relatively large hop for the data misfit. Moving from regularization parameter $\beta = 5.7 \times 10^{-4}$ to $\beta = 5.1 \times 10^{-4}$, the data misfit hops from 5.37 to 1.14, and from 1.14 to 0.88 ($=\epsilon_{\text{RMS}}^2$) with  $\beta = 4.6 \times 10^{-4}$. We select the optimal regularization parameter $\beta$ closest to $\epsilon_{\text{RMS}}$ equal to one (and thus $\beta = 4.6 \times 10^{-4}$). In this case, this is $\epsilon_{\text{RMS}} = 0.94$, which is slightly overfitted.  The inversion model in Figure \ref{fig:V_11} nicely demonstrates the blocky structure, as expected from the db1-db1 combination. We check if $\alpha$ is a proper choice by checking the parameter $\gamma$. With $\gamma = 1$, there are significant variations in both directions, yet it may seem that there is too much variation in the lateral direction. For example, for sounding number 30 in Figure \ref{fig:V_11}, the typical blocky structure is prominent, as expected from an  1D regularization scheme with a good candidate for regularization parameter $\beta$. For the lateral variation, for example at a depth of $\pm 15$ m, the inversion model agrees with the true model. The issue here is that this inversion model does not exhibit the minimum structure, as we would expect from a blocky inversion. The transitions are relatively smooth and thus costly (in terms of lateral model misfit $\phi_{m,\leftrightarrow}$). Increasing the relative cost of the lateral model misfit functional or thus $\alpha$ would overcome this issue. This is done manually by setting the $\gamma$-parameter. \\

In the case of a suboptimal $\gamma = 1$ inversion model, a sweep over $\gamma$ is warranted. An example of a sweep is shown in Figure \ref{fig:sweep}, where the inversion models for $\gamma = 0.1, 0.5, 5, 15, 50 $ and 100 are shown (for $\gamma = 1$, refer back to Figure \ref{fig:V_11}). For $\gamma = 0.1$ for example, at first sight, a minimum-structure block from $-13$ to $-27$ m depth and 0 to approx. 800 m in the lateral orientation is observed. However, there is quite some heterogeneity in that block that is a result of overfitting. For $\gamma = 100$, we observe excessive minimal structure in the lateral orientation, seen as elongated building blocks. The $\gamma = 5$ inversion model seems to be a good candidate as it seems to be less overfitted than the $\gamma=1$ inversion model. Its data misfit is 1.04, which is close to one and the inversion model itself is blocky in both vertical and lateral directions. Below each inversion model, so-called `error profiles' are shown, which show the noise weighted root mean-squared per sounding $\epsilon_{\text{RMS}}^s$, which allow for a more thorough quality control. There is more variation in the error profiles if sharper transitions are recovered. However, a multidimensional forward modelling should be performed to use error profiles for quantitatively assissing the quality of such a quasi-2D inversion, as the 1D assumption is possibly violated \citet{deleersnyder2022novel}. \\

Once the user-defined $\gamma$-parameter has been picked, the manifold of the inverse problems can be constructed, where all the potential interesting combinations of the wavelet basis functions can be used, such as presented in Figure \ref{fig:VI_2D}.

\begin{table}
	\centering
	\caption{Estimated parameter $\tilde{\alpha}$ and model discrepancy $|| \vb{M}   - \vb{M}^{\text{true}} ||$ for each wavelet-wavelet combination of Figure \ref{fig:VI_2D} and the model discrepancy for the inversion model from the LCI method of Figure \ref{fig:lci}{ with constraints $c$ of 1.1.} }
	\label{tab}
	\begin{tabular}{l | l | l }
		Wavelet combination &  $\tilde{\alpha}$  & $|| \vb{M}   - \vb{M}^{\text{true}} ||$ \\
		\hline
		db1-db1 & 0,500 & 4.249\\
		db1-db2 & 2,82 & 3.955\\
		db1-db5 & 58,1 & 3.063\\
		db2-db3 & 0,750 & 3.417\\
		db3-db2 & 0,0450 & 3.012\\
		db3-db8 & 5,45 & 2.896\\
		db5-db1 & 0,00721 & 3.255\\
		db6-db5 & 0,069 & 3.553\\
		db6-db9 & 0,600 & 3.221\\
		LCI (c=1.1) & / & 2.944 \\
		\hline
	\end{tabular}
\end{table}

In Figures \ref{fig:VI_2D}A-C, the inversion models with db1 in the vertical orientation are shown. Figure \ref{fig:VI_2D}B has larger amplitudes than Figure \ref{fig:VI_2D}A and sharp transitions in the lateral orientation. Figure \ref{fig:VI_2D}C also has larger amplitudes, but the lateral transitions are smoother. The $\epsilon_{\text{RMS}}$ is rather low, but the inversion model with a slightly larger regularization parameter $\beta$ is severely under fitted and lies further from $\epsilon_{\text{RMS}}$ equal to one. An alternative for the under fitted inversion model would be to look at inversion models with similar properties, such as db1-db6, as they potentially have an inversion model with a root-mean-squared error $\epsilon_{\text{RMS}}$ closer to one. However, db1-db$n$ inversion models turn out to be prone to local minima (something that is infrequently observed with other combinations).\\

\begin{figure*}
	\centering
	\begin{tabular}{l}
		A. \\
		\includegraphics[width=0.80\linewidth]{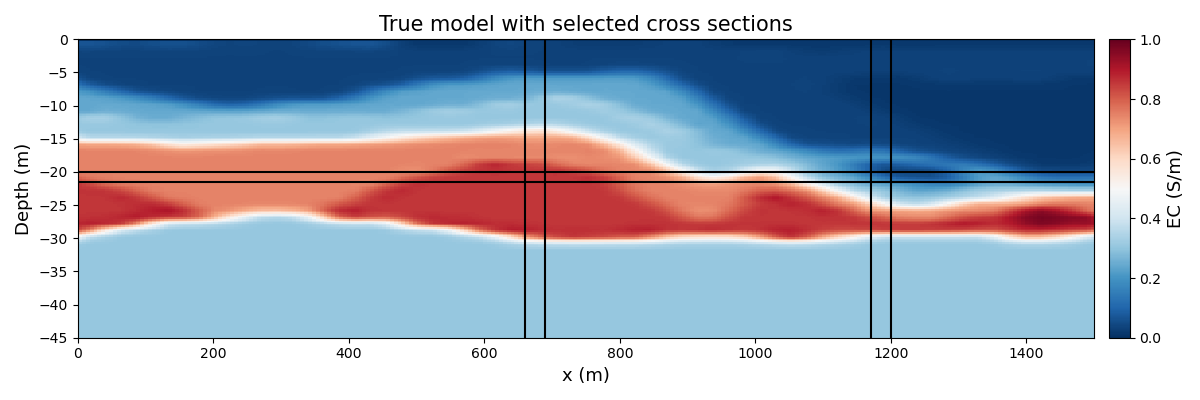}  \\  

		B. \hspace{0.4\linewidth} C. \\
		\includegraphics[width=0.4\linewidth]{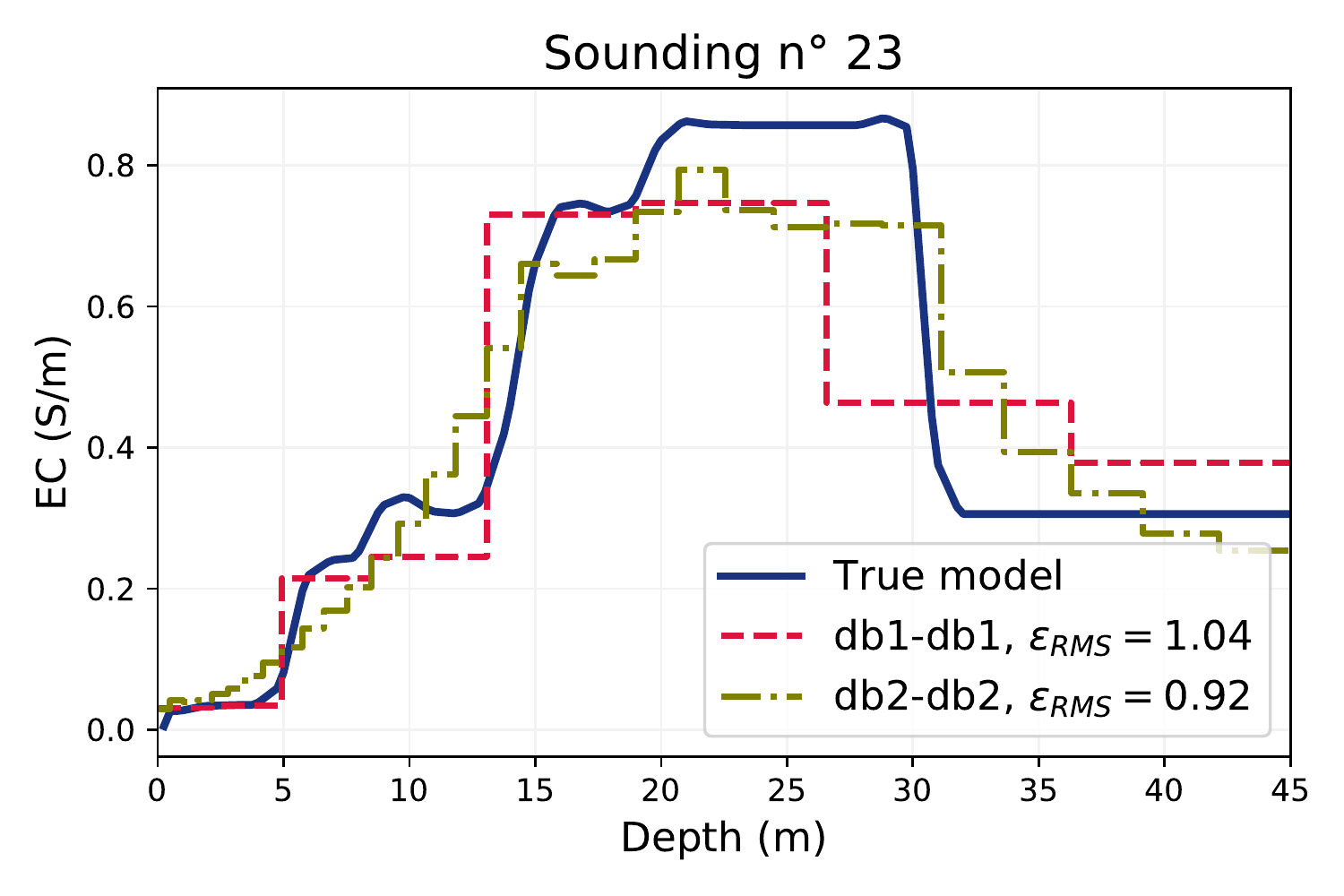}  \includegraphics[width=0.4\linewidth]{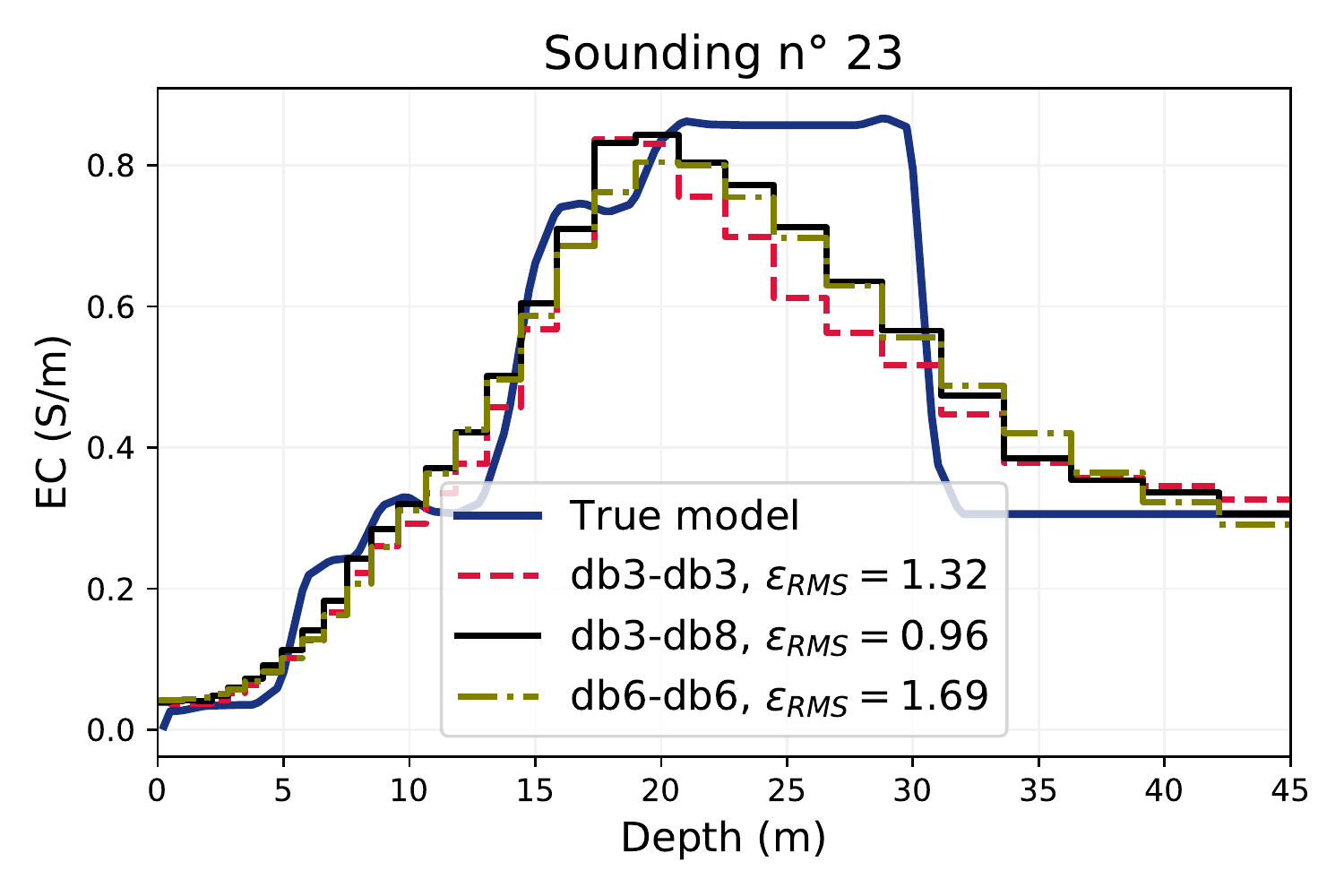}  \\ 

		D. \hspace{0.4\linewidth} E. \\
		\includegraphics[width=0.4\linewidth]{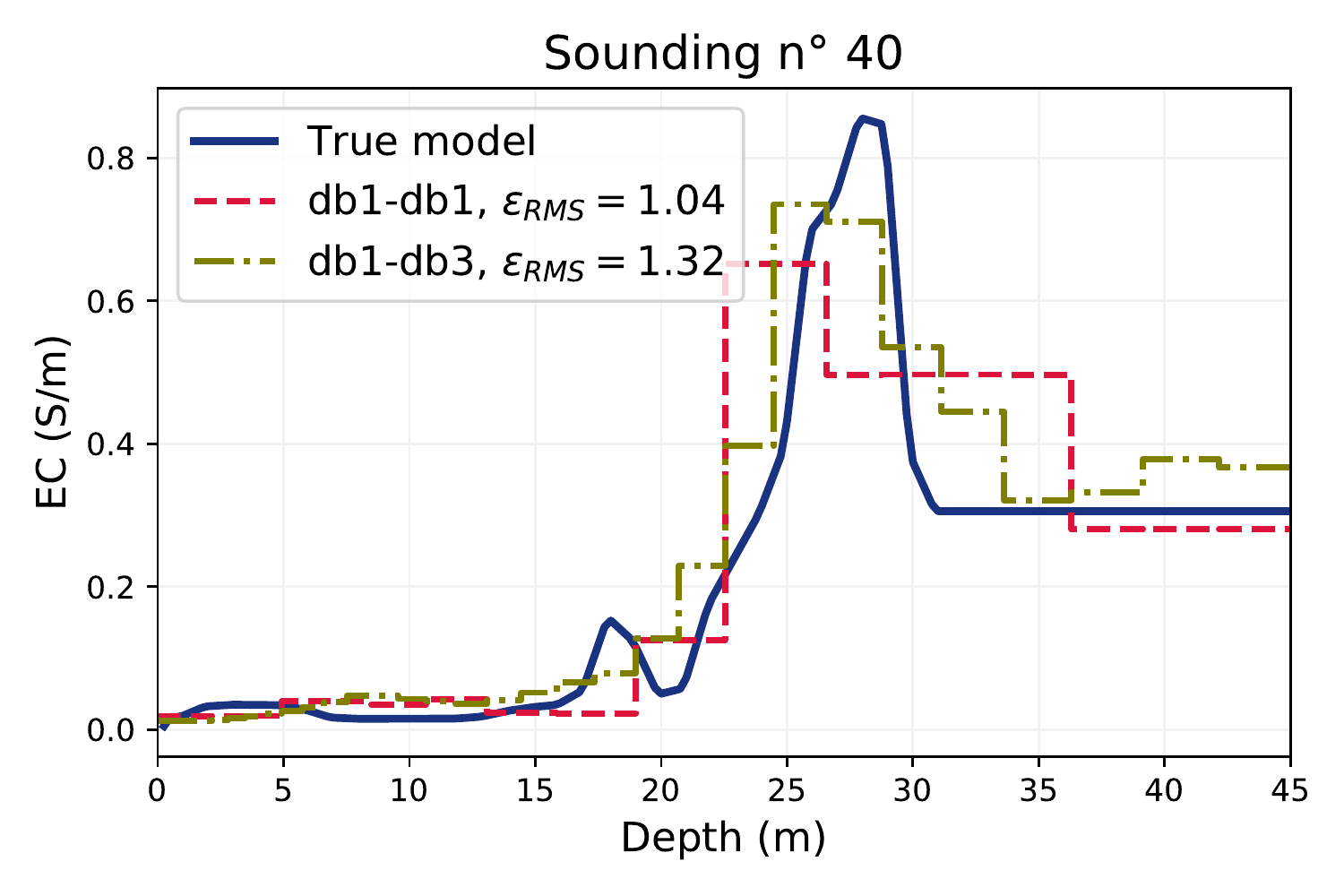}  \includegraphics[width=0.4\linewidth]{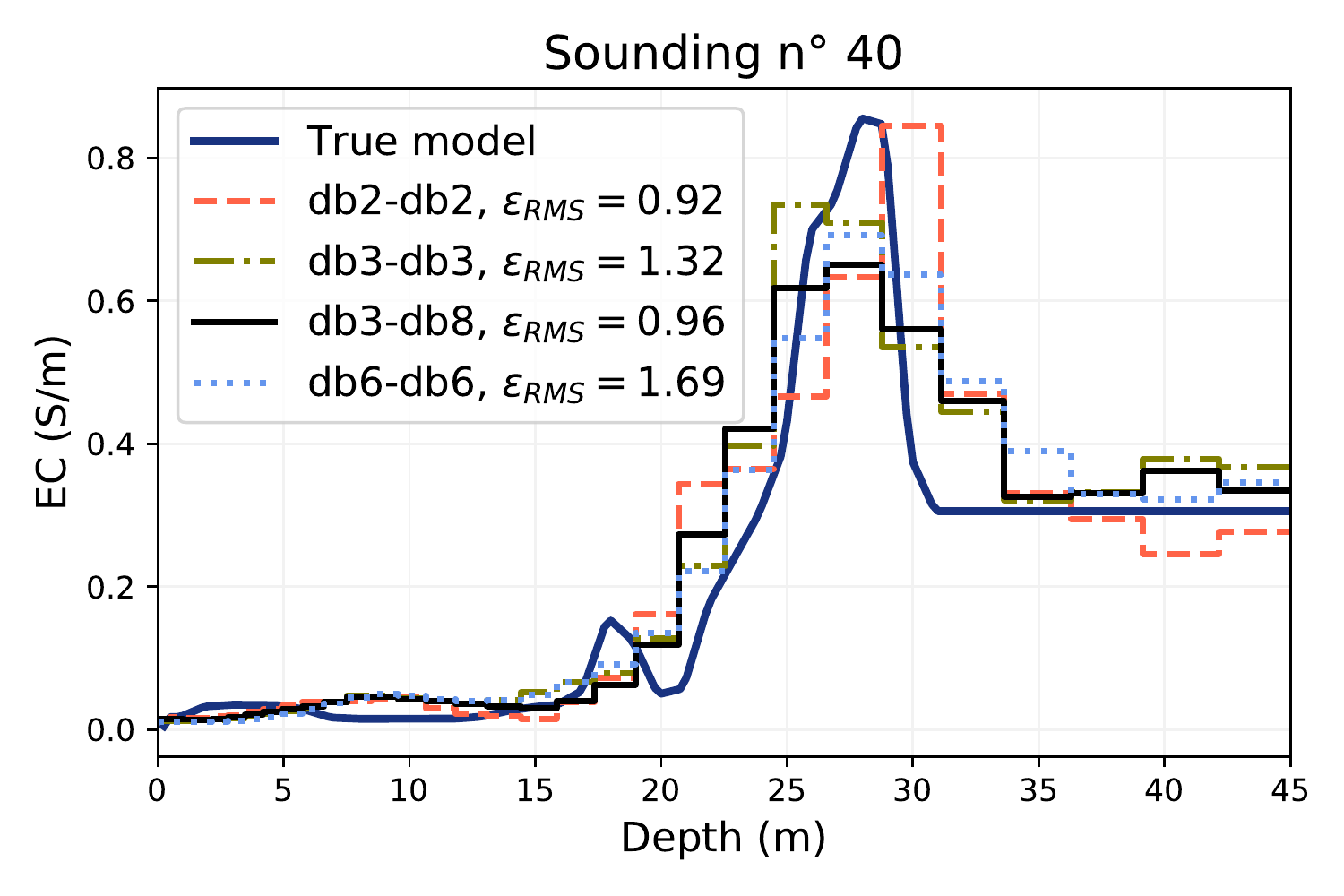}   \\ 

		F. \\
		\includegraphics[width=0.80\linewidth]{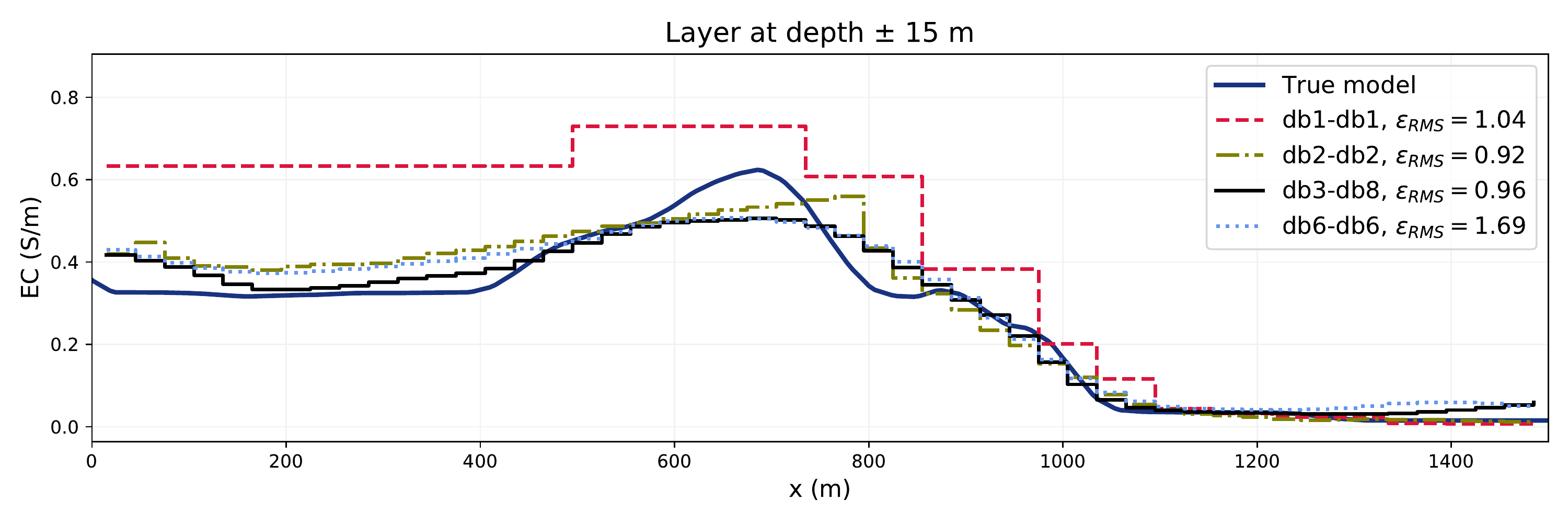}  
	\end{tabular}
	\caption{Selected conductivity profiles for the synthetic model. Different inversion models uncover different characteristics of the wavelet basis function (e.g. blocky, smooth, intermediate).} 
	\label{fig:VI_cross}
\end{figure*}

 Figure \ref{fig:VI_2D}A proves that our regularization scheme can recover blocky structures (which is an appealing feature in other cases), however, our synthetic model is not that blocky and thus this inversion model is not expected. The db6-db9 inversion model in Figure \ref{fig:VI_2D}I is the other limiting case and is comparable with smoothing regularization. The problem with smoothness regularization is that the minimal and maximal electrical conductivity values tend to be over- and underestimated, respectively. Here, this is also the case, see for example Figure \ref{fig:VI_2D}I, where the maximum value of the recovered electrical conductivity is 0.834 S/m while the maximum is 0.97 S/m in the true model. All other inversion models exhibit intermediate results, e.g., db2-db3 can recover a large peak in the vertical orientation. The db3-db2 inversion model, on the other hand, which has an identical data misfit, generates a smoother peak (see Figure \ref{fig:VI_cross}E), which is for this inversion model more favourable. In our synthetic data case within the saltwater intrusion context, we know that the transition from fresh to saltwater is sharp, but not blocky. For example, db3 exhibits this feature. In the lateral orientation, however, relatively smooth profiles are expected. Thus db3-db8 in Figures \ref{fig:VI_2D}F and \ref{fig:VI_cross}C, E, F which also fits the data quite well ($\epsilon_{\text{RMS}}$ = 0.96) would be a good candidate. It can both recover the electrical conductivities and the global structure of the inversion model quite well, such as the rather subtle saltwater lens at 1000 m. (Note that a finer discretization, which comes with a higher computational cost, would yield better estimations of the peaks in Figure \ref{fig:VI_cross}.) Note that the error profiles with $\epsilon_{\text{RMS}}^s$ for inversion models with sharp lateral transitions vary more than those with smooth transitions. The error is closer to one for $x$ larger than 1200 m for sharp transitions than for smoother inversion models.\\

\begin{figure}
	\centering 
	\includegraphics[width=0.5\textwidth]{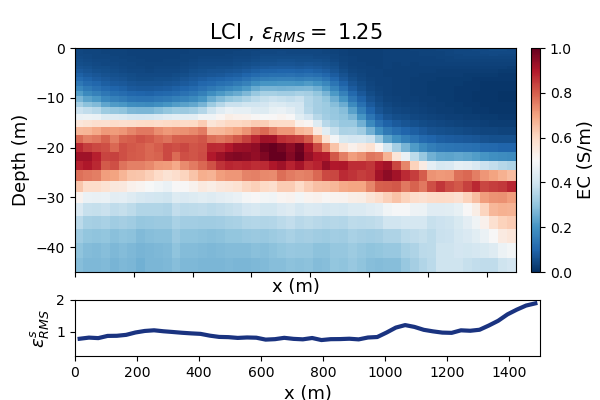}
	\caption{Laterally Constrained Inversion on the synthetic model with 1.1 as both lateral and horizontal constraints.} 
	\label{fig:lci}
\end{figure}

We have discussed the flexibility of the method based on features that can be recognized in the inversion models. We now assess the performance of our method and compare our inversion models with respect to the true model. In this analysis, we add a common method in AEM inversion to check whether our proposed method is capable of obtaining comparable results. In AEM inversion, Laterally Constrained Inversion is quite standard and the implementation of \citet{auken2015overview} was used to generate the inversion model in Figure \ref{fig:lci}, using constraints of 1.1. The performance of the inversion models is measured by the model discrepancy $|| \vb{M}   - \vb{M}^{\text{true}} ||$. The values are presented in Table \ref{tab}. The results were obtained by projecting the true model to the model discretization in which the inversion was performed, as shown in Figure \ref{fig:true_model_inversion_discretization}. The results of Table \ref{tab} confirm that db3-db8 is a good choice for the type or minimum structure, as it depicts the lowest model discrepancy of all inversion. This wavelet-wavelet combination is able to recover the true model as well as the LCI method (similar model discrepancy), although the latter tends to retrieve more lateral variations. The combinations db3-db2 and db1-db5 are two satisfactory alternatives. Note the large model discrepancy for db1-db1 combination, which was to be expected. The db1-db1 inversion model is not a good match with the true model, but it fits the data ( $\epsilon_{\text{RMS}}$ = 1.04), which again illustrates the non-uniqueness of the inverse problem. Accordingly, db1-db1 is  not a good choice for this application, but can be attractive for e.g., archaeological applications. Interestingly, the error profile of the LCI inversion model in Figure \ref{fig:lci} also shows larger misfits for $x$ larger than 1200 m. As a quality control, the methods perform quite similar. The db3-db8 inversion model is fitting each sounding to $\epsilon_{\text{RMS}}^s = 1$ more consistenly than this LCI inversion model.

\subsection{Field data case}
\label{sec:fielddata}

In this section, we apply the flexible quasi-2D wavelet-based inversion scheme on the airborne data of Belgian's current salinization map of vulnerable regions in Flanders \citep{vlaanderentopsoil}. The salinization map shows the interface between salt and freshwater and serves as decision tool for policymakers to assess local measures to increase freshwater availability, which may be hindered by future effects of climate change. The use of quasi-2D inversion is justified, as the distance between the soundings along the line of flight is approximately 30 m, while much larger between the line of flights (between 250-275m).\\

\begin{figure*}
	\begin{tabular}{l}
%
%
	\includegraphics[width=\linewidth]{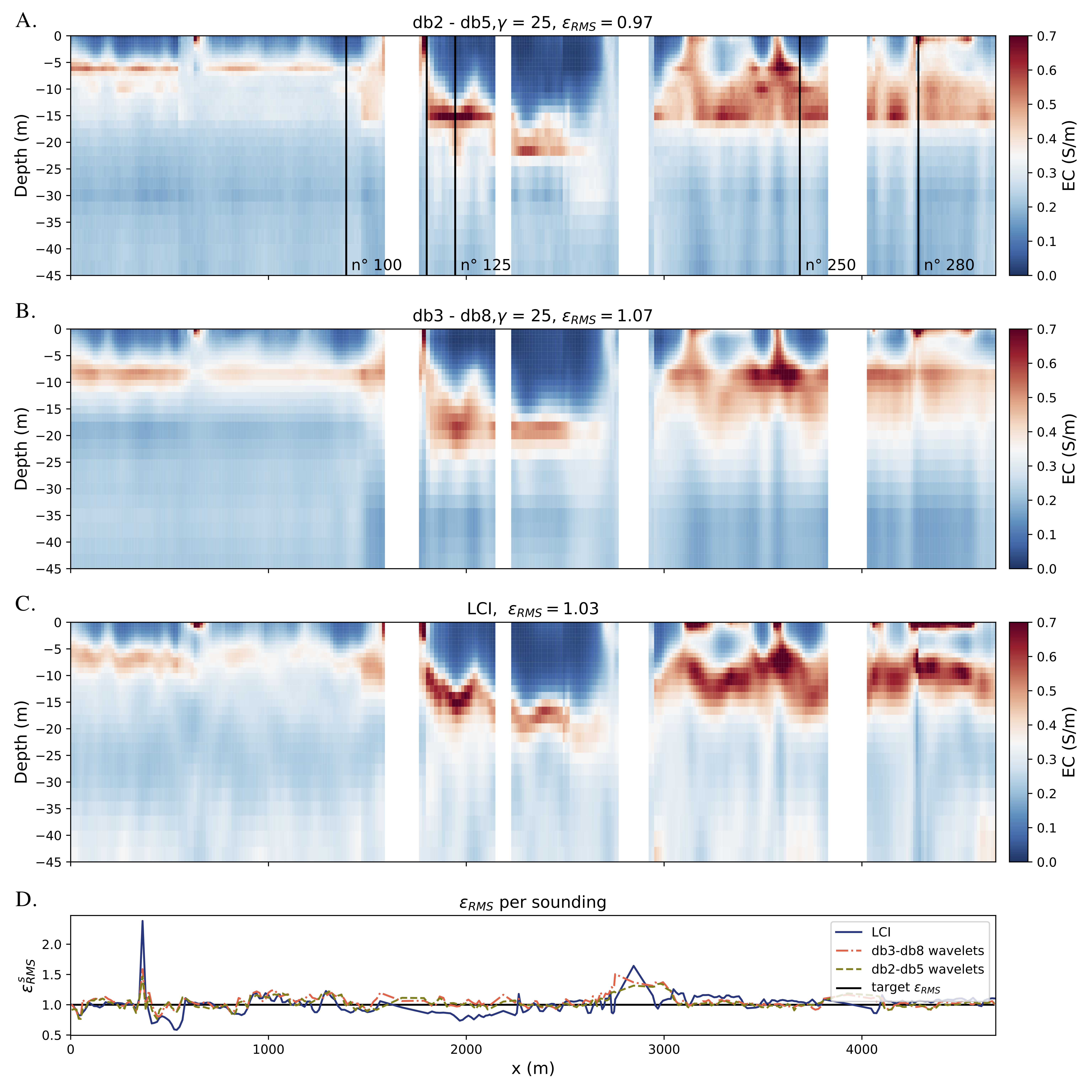}\\

	\end{tabular}
	\caption{Inversion models of the Yzercreek profile with db2-db5 regularization (A.), db3-db8 regularization (B.), and (C.) Laterally Constrained Inversion (LCI) (constraint between horizontally neighbouring cells was $c = 1.3$ and vertically $c=2$). Error profiles for each inversion model are shown in D.} 
	\label{fig:real_data}
\end{figure*}

\begin{figure*}
	\includegraphics[width=\linewidth]{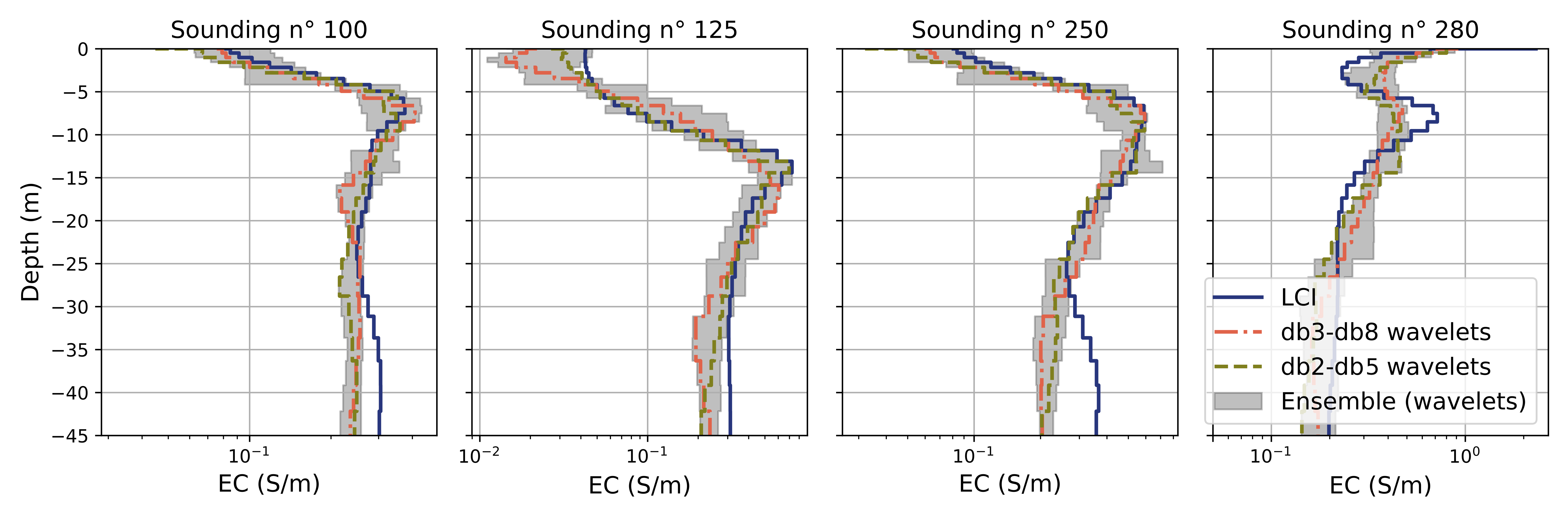}
	\caption{Comparison between two wavelet-wavelet regularization inversion models and the Laterally Constrained Inversion result. The ensemble of acceptable wavelet-wavelet regularization inversion models indicates the non-unicity of the recovered model.}
	\label{fig:ensemble}
\end{figure*}

We perform an inversion on a flight line from Flanders' salinization map TDEM sub dataset with 300 soundings. The data were acquired using a SkyTEM system, as described in Section \ref{sec:results}, in 2017 as part of the Topsoil project \citep{vlaanderentopsoil}. The flight line crosses the Yzercreek, where creek ridge deposits occur which can be locally thick. \citet{zeuwts1991hydrogeologie} reports a thick freshwater lens (TDS $<$ 1000 mg/l), while other creek ridge deposits are thinner. At some locations, solely brackish and saline groundwater can be found under the shallow subsurface consisting of a thin cover of heavy clay resting on peat and lying between higher creek ridge soils.\\

The optimal $\gamma$-parameter was picked using the same method as in the synthetic data case based on the db1-db1 plot. The optimal $\gamma$-parameter was set to 25. All inversion models with each wavelet-wavelet combination were generated. However, we use the conclusion from the synthetic field case, which is similar to this setting, that db3-db8 is a good choice for the wavelet basis functions. For completeness, we add the db2-db5 inversion model, which results in sharper transitions in each orientation, because the true sharpness of especially the vertical orientation is not known. Those inversion models are shown in Figures \ref{fig:real_data}A and B. We can again compare this result with the LCI method, presented in Figure \ref{fig:real_data}C. The constraints for the LCI method are taken from \citet{vlaanderentopsoil} (constraints between vertical neighbouring cells is 2, while constraints between horizontal neighbouring cells is 1.3).\\
	
We recognize similar trends between the inversion models. A difference with the LCI method is that the wavelet-based method can identify the anomalies with a higher precision, such as at the surface at 625 m and 4100 m, because the highly conductive values are not smeared out due to the smoothness constraint. This is a property of wavelet-based inversion that was found in earlier work, where we have found that wavelet-based inversion can recover high amplitude anomalies in globally smooth profiles. At 5 and 10 m depth the inversion model with db3-db8 looks slightly smeared out compared to the LCI method, while the db2-db5 inversion model is sharper and recovers higher amplitudes.\\
	
Four vertical profiles of each inversion are shown in Figure \ref{fig:ensemble}, together with the ensemble of recovered wavelet-wavelet models that have converged to at least $\epsilon_{\text{RMS}} < 1.2$. The peaks in the conductivity profiles are almost exactly at the same depths and have similar electrical conductivity values, except for sounding 280, where we observe a discrepancy. Viewing the results as an ensemble provides an indication of the non-unicity or the range of acceptable EC-values and their variability. For sounding 100, we have a large variability in the first 5 m of the profile, but a more unequivocal result around the transition at 5 m. The db3-db8 model follows the LCI model well, while the db2-db5 has a lower EC value that starts and peaks at a slightly higher EC value, which is to be expected for a sharper inversion. The blocky transition can be well recognized in the ensemble at approximately 4 m depth. From a depth of 25 m, the LCI model is not in agreement with the results from the ensemble. For sounding 125, the non-uniqueness is more present. Note that the slightly smoother db3-db8 still has a lower EC value in the first meters than the db2-db8. The EC value of the LCI model in the first few meters is at the upper limit of the ensemble. Interestingly, the range of variation in the inversion results varies throughout the soundings, as sounding 250 indicates much less variation. Sounding 280 shows more variability at larger depths than the other soundings. This can be explained by a reduced sensitivity due to the `shielding' of the highly conductive artefact right below the surface. The variation right below the surface is an indication of the non-unicity of the inverse problem. Observe that the LCI result for the EC-value at -2 m for sounding 280 is within the wavelet-wavelet ensemble, while the peak near -7.5 m is not.\\

From the analysis of the noise-weighted error per sounding $\epsilon_{\text{RMS}}^s$ in Figure \ref{fig:real_data}D, we infer that our method succeeds in fitting the soundings individually. In this sense, the two wavelet-wavelet inversion models shown are equivalent. That the most significant discrepancies (at approx. 400 m and 2800-3000 m) occur at the same locations (as with the LCI method) makes this claim even more convincing.
\section{Discussion}
A major strength of the presented method over conventional deterministic methods is that it can adapt the sharpness differently in both orientations by simply tuning the parameters, that is the number of vanishing moments of the wavelet basis function used in each orientation. The flexibility is limited to solely a discrete set of wavelet basis functions. However, we were already able to recover multiple types of minimum structure with only one wavelet basis family (Daubechies wavelets), such as blocky, smooth and intermediate sharpness. Other wavelet basis functions in combination with the notion of scale-dependency are yet unexplored. Moreover, the number of model parameters, i.e. the discretization of the inversion grid, can also be varied, as this also influences the final sharpness of the result, as with e.g. Tikhonov regularization.\\

It is beyond the scope of this study to examine the `optimal' choice of (multiple) basis function(s), as the result is likely dependent on the nature of the true model. The flexibility of the proposed inversion scheme allows for more complex inversion strategies and can produce multiple subsurface realizations that fit the data equally well, yet `simpler' in Occam's sense. It will be part of future work to define criteria for finding the 'best' wavelet basis. For now, the geoscientist will need to trust expert knowledge, guiding geological data, and acknowledge that the only true criterion is based on the data misfit. The ensemble of inversion models allows revealing different features within the same framework of wavelet-based inversion. If more data or prior knowledge is available, a different inversion model from the wavelet-wavelet ensemble may become more suited. Also note that our choice to go for the db3-db8 wavelet-wavelet combination was made on its global performance, but this overall good performance is not necessarily true at each sounding location.\\

The optimization problem in the model domain does not significantly complicate our previously developed regularization scheme in 1D \citep{deleersnyder2021inversion}, which was originally developed to be optimized in the wavelet domain. The persistent presence of local minimal in the db1-db$n$ for $n>3$ may be ascribed to that change, as is the extra complexity of the model misfit as the sum of two separate model misfits. We currently do not have a substantiated cause. The first solution to overcome this problem would be to start from another random inversion model until an acceptable $\epsilon_{\text{RMS}}$ after convergence is obtained, though this did not resolve this issue. Our proposed solution is to use the e.g. db3-db5 outcome as starting model for db1-db5. This will sharpen the transitions of the inversion model along the vertical orientation, while still fitting the data.\\

Alternatives to this flexible method are e.g. \citet{klose2022laterally} who present a flexible approach via the focusing parameter of the minimum gradient support functional for frequency-domain electromagnetic induction surveys in a bi-directional (quasi-2D inversion models) setting. This approach also uses two model misfits, one for each orientation, and acknowledges the multi-solution strategy for the non-uniqueness of the problem. Given the different setting, the performance of the methods cannot be compared. Other alternatives exist that utilize 2D wavelet-based regularization. A growing body of literature focuses on (multi)directionality: how do we recover diagonal features? As previously mentioned, \citet{nittinger2016inversion} therefore specifically uses a 2D wavelet with six directions. In recent work \citep{su2021sparse}, the shearlet transform is utilised. This is an extension of the wavelet transform that is supplemented with some notion of directionality. This gives rise to an even higher redundancy: A 64$\times$64 inversion model requires a representation of $606~208$ shearlet coefficients on which the sparsity condition is imposed (at a priori unknown indices). Our findings are that we do not experience any problems with directionality (see for example Section \ref{sec:fielddata}, which demonstrate that features are recovered with diagonal characteristics, illustrating that a shearlet transform, as introduced in the Introduction, is not necessary in this context.), certainly not so that they complicate the geological interpretation. For the db1-db1 case, diagonal features tend not to be recovered well. This may be exploited in settings where bi-directional blocky structures are expected and thus still clearly has an advantage.

Another advantage of our approach is that no square inversion models are required, which is certainly welcome for AEM inversion where the number of soundings along a profile is typically much bigger than the number of model parameters in the vertical orientation. In other wavelet-based methods, where for example an inversion model with a shape of 30 by 64 would be ideal, it should be extended to a 64$\times$64 inversion model in order to apply the 2D discrete wavelet transform, which in turn affects the computation time of the Jacobian of the data-fitting term.\\

The scale-dependent wavelet-based model misfit can easily be applied to other geophysical inverse problems. The calibration procedure proposed in Section \ref{sec:calibration} cannot be straightforwardly applied to other geophysical methods. Adjustments would probably be needed, but the underlying principle is useful. For example, see also another calibration procedure for surface wave dispersion, recently proposed by \citet{guillemoteau2022sparse}, where the relative regularization parameter is manually tuned during the iteration process, by comparing the lateral smoothness to the variation in the dispersion curves. \citet{hermans2016covariance} uses first an isotropic smooth inversion to estimate the ratio and to apply to subsequent covariance-based inversion.\\

The advantage of such a calibration procedure cannot be minimised. As mentioned in Section \ref{sec:calibration}, the use of different model misfits along each orientation, with potentially different number of model parameters or wavelet basis, has a large impact on the absolute value of the misfit, independent of the complexity of the inversion model along that direction. Hence, the relative regularization parameter $\alpha$ plays an important role in correcting the relative importance of the two misfits. In Section \ref{sec:synthetic} we found (Table \ref{tab}) that there is quite some variability in $\tilde{\alpha}$. This shows on the one hand the need for such a calibration technique but also its effectiveness. It relaxes the need for an extensive optimization for $\alpha$, which saves a significant amount of computational energy. The method is not fully automatic yet, hence the parameter $\gamma$ that allows the user to fine-tune the result and use geological expertise or prior knowledge. We have demonstrated that this parameter has to be tuned only once and then can then be applied to the whole ensemble, which is again advantageous in terms of computational cost. Note that the calibration procedure will be less effective when the airborne loops undergo altitude variations, as this leads to variability in the TDEM data. In Section \ref{sec:fielddata}, where we discuss the field data case, the calibration strategy could still be applied, but this caused a larger $\gamma$ parameter. A height-correction could be developed in future work, making the $\gamma$ parameter potentially obsolete.\\

A disadvantage of this proposed inversion scheme is that it uses a 1D forward model, because multidimensional forward modeling for time-domain EM problems is computationally demanding. The method for looking at the error profiles (the error per sounding) is therefore not entirely quantitative: the higher small-scale variability in the error profiles for sharper inversion results are an effect of the 1D forward modelling. For 2.5D forward modelling, the error profiles would be much smoother and therefore more `equivalent'. An possible solution is to apply an image appraisal tool for imperfect forward modelling after the quasi-2D inversion. In \citet{deleersnyder2022novel}, a computational efficient method is proposed that can be used to assess whether multidimensionality issues are present, meaning that there are datapoints that fit the observed data well with a 1D forward model, but not with a 2.5D or 3D forward model. With this method it can be decided whether multidimensional modelling is needed in a certain context or not. This is especially important for inversion models obtained with a lateral wavelet basis function with few vanishing moments.

\section{Conclusion}
The multidimensional, scale-dependent wavelet-based inversion scheme is an alternative scheme that is more flexible than smoothness or blocky inversion and can easily be combined with existing frameworks of deterministic inversion (gradient-based optimization methods, the discrepancy principle for optimal regularization parameter). It can recover inversion models with tunable sharpness and limiting cases (smoothness and blocky inversion). The regularization term uses the wavelet transform of an inversion model in combination with the Ekblom measure to estimate the complexity of an inversion model in Occam's sense. The choice of the wavelet basis function underlying the wavelet transform allows for defining the features in the final inversion model. \\

Our approach is different from other wavelet-based regularization schemes, as this regularization method allows for choosing a different wavelet basis function for each orientation and therefore, for example, obtaining sharp results along with the vertical orientation and smoother results along with the lateral orientation, which is a desirable feature for many geological contexts, such as saltwater intrusion. The difficulty of pinning down the relative regularization parameter (vertical vs. horizontal complexity) is resolved with a calibration step, which uses the variability in the EM data and the complexity of one (or more) random sounding(s). We have shown that the calibration step is at least a good starting point for the optimization of that relative regularization parameter. Optimizing further this parameter requires specific investigations. \\

Depending on the availability of prior knowledge, different interpretation approaches can be used. If no prior knowledge is available, the high flexibility of the method can be exploited to easily generate a set of representations of the inverse problem, each highlighting different features, while fitting the data. Common features in the set of inversion models are an indication that the feature is in the geophysical data. When borehole loggings are available, the profile could be correlated with the colocated data to calibrate the sharpness along with the vertical orientation. In our example, we have used an existing profile to generate synthetic data similarly to our real field data. By analysing the ensemble of inversion models and comparing it to the original profile, we have determined that the db3 wavelet (relatively sharp) along the vertical orientation with the db8 wavelet (smooth) along the lateral orientation is a suitable candidate for the field data inversion.\\

We have demonstrated the potential of the flexible regularization for airborne time-domain EM data, where a different and appropriate sharpness was used in each orientation. However, the method could be equally applied to any other geophysical method. It can be straightforwardly extended to quasi-3D inversion or it can be used with full 2.5D or 3D forward models.

\section*{ACKNOWLEDGEMENTS}
The authors thank VMM (Flanders Environment Agency) for making the data behind Flanders' salinization map available. The research leading to these results has received funding from FWO (Fund for Scientific Research, Flanders, grant 1113020N and 1113022N), the Flemish Institute for the Sea (VLIZ) Brilliant Marine Research Idea 2022 and the King Baudouin Foundation Ernest du Bois prize 2022. The resources and services used in this work were provided by the VSC (Flemish Supercomputer Center), funded by the Research Foundation - Flanders (FWO) and the Flemish Government. We thank the Aarhus HydroGeophysics group for granting us a more extensive than standard academic license for the AarhusInv software, underlying the Laterally Constrained Inversion results. We also acknowledge Julien Guillemoteau and two anonymous reviewers for their thoughtful and valuable comments on this paper.

\section*{DATA AVAILABILITY}
The 2.5D forward data from the synthetic data case are provided in the supplementary materials.  The field data that support the findings of the field data case are available from the corresponding author, Wouter Deleersnyder, upon reasonable request.

\bibliographystyle{gji}

\begin{thebibliography}{52}
	\expandafter\ifx\csname natexlab\endcsname\relax\def\natexlab#1{#1}\fi
	
	\bibitem[Anderson(1979)]{anderson1979numerical}
	Anderson, W.~L., 1979.
	\newblock Numerical integration of related hankel transforms of orders 0 and 1
	by adaptive digital filtering, {\it Geophysics\/}, {\bf 44}(7), 1287--1305.
	
	\bibitem[Auken \& Christiansen(2004)]{auken2004layered}
	Auken, E. \& Christiansen, A.~V., 2004.
	\newblock Layered and laterally constrained 2d inversion of resistivity data,
	{\it Geophysics\/}, {\bf 69}(3), 752--761.
	
	\bibitem[Auken et~al.(2015)Auken, Christiansen, Kirkegaard, Fiandaca, Schamper,
	Behroozmand, Binley, Nielsen, Effers{\o}, Christensen,
	et~al.]{auken2015overview}
	Auken, E., Christiansen, A.~V., Kirkegaard, C., Fiandaca, G., Schamper, C.,
	Behroozmand, A.~A., Binley, A., Nielsen, E., Effers{\o}, F., Christensen,
	N.~B., et~al., 2015.
	\newblock An overview of a highly versatile forward and stable inverse
	algorithm for airborne, ground-based and borehole electromagnetic and
	electric data, {\it Exploration Geophysics\/}, {\bf 46}(3), 223--235.
	
	\bibitem[Auken et~al.(2017)Auken, Boesen, \& Christiansen]{auken2017review}
	Auken, E., Boesen, T., \& Christiansen, A.~V., 2017.
	\newblock A review of airborne electromagnetic methods with focus on
	geotechnical and hydrological applications from 2007 to 2017, {\it Advances
		in Geophysics\/}, {\bf 58}, 47--93.
	
	\bibitem[Christensen(2016)]{christensen2016strictly}
	Christensen, N.~B., 2016.
	\newblock Strictly horizontal lateral parameter correlation for 1d inverse
	modelling of large datasets, {\it Near Surface Geophysics\/}, {\bf 14}(5),
	403--412.
	
	\bibitem[Cockett et~al.(2015)Cockett, Kang, Heagy, Pidlisecky, \&
	Oldenburg]{cockett2015simpeg}
	Cockett, R., Kang, S., Heagy, L.~J., Pidlisecky, A., \& Oldenburg, D.~W., 2015.
	\newblock {SimPEG}: An open source framework for simulation and gradient based
	parameter estimation in geophysical applications, {\it Computers {\&}
		Geosciences\/}, {\bf 85}, 142--154.
	
	\bibitem[Constable et~al.(1987)Constable, Parker, \&
	Constable]{constable1987occam}
	Constable, S.~C., Parker, R.~L., \& Constable, C.~G., 1987.
	\newblock Occam’s inversion: A practical algorithm for generating smooth
	models from electromagnetic sounding data, {\it Geophysics\/}, {\bf 52}(3),
	289--300.
	
	\bibitem[Cox et~al.(2010)Cox, Wilson, \& Zhdanov]{cox20103d}
	Cox, L.~H., Wilson, G.~A., \& Zhdanov, M.~S., 2010.
	\newblock 3d inversion of airborne electromagnetic data using a moving
	footprint, {\it Exploration Geophysics\/}, {\bf 41}(4), 250--259.
	
	\bibitem[Daubechies(1988)]{daubechies1988orthonormal}
	Daubechies, I., 1988.
	\newblock Orthonormal bases of compactly supported wavelets, {\it
		Communications on pure and applied mathematics\/}, {\bf 41}(7), 909--996.
	
	\bibitem[Daubechies et~al.(2004)Daubechies, Defrise, \&
	De~Mol]{daubechies2004iterative}
	Daubechies, I., Defrise, M., \& De~Mol, C., 2004.
	\newblock An iterative thresholding algorithm for linear inverse problems with
	a sparsity constraint, {\it Communications on Pure and Applied Mathematics: A
		Journal Issued by the Courant Institute of Mathematical Sciences\/}, {\bf
		57}(11), 1413--1457.
	
	\bibitem[Deleersnyder et~al.(2021)Deleersnyder, Maveau, Hermans, \&
	Dudal]{deleersnyder2021inversion}
	Deleersnyder, W., Maveau, B., Hermans, T., \& Dudal, D., 2021.
	\newblock Inversion of electromagnetic induction data using a novel
	wavelet-based and scale-dependent regularization term, {\it Geophysical
		Journal International\/}, {\bf 226}(3), 1715--1729.
	
	\bibitem[Deleersnyder et~al.(2022{\natexlab{a}})Deleersnyder, Dudal, \&
	Hermans]{deleersnyder2022novel}
	Deleersnyder, W., Dudal, D., \& Hermans, T., 2022{\natexlab{a}}.
	\newblock Novel airborne em image appraisal tool for imperfect forward
	modeling, {\it Remote Sensing\/}, {\bf 14}(22), 5757.
	
	\bibitem[Deleersnyder et~al.(2022{\natexlab{b}})Deleersnyder, Dudal, Maveau, \&
	Paepen]{deleersnyder2022determining}
	Deleersnyder, W., Dudal, D., Maveau, B., \& Paepen, M., 2022{\natexlab{b}}.
	\newblock Determining the optimal focusing parameter in sparse promoting
	inversions of emi surveys, {\it arXiv preprint arXiv:2211.12552\/}.
	
	\bibitem[Delsman et~al.(2019)Delsman, van Baaren, Vermaas, Karaoulis, Bootsma,
	de~Louw, Pauw, Oude~Essink, Dabekaussen, Van~Camp, Walraevens, Vandenbohede,
	Teilmann, \& Thofte]{vlaanderentopsoil}
	Delsman, J., van Baaren, E., Vermaas, T., Karaoulis, M., Bootsma, H., de~Louw,
	P., Pauw, P., Oude~Essink, G., Dabekaussen, W., Van~Camp, M., Walraevens, K.,
	Vandenbohede, A., Teilmann, R., \& Thofte, S., 2019.
	\newblock Topsoil airborne em kartering van zoet en zout grondwater in
	vlaanderen, Tech. rep., VMM.
	
	\bibitem[Donoho(2006)]{donoho2006most}
	Donoho, D.~L., 2006.
	\newblock For most large underdetermined systems of linear equations the
	minimal l1-norm solution is also the sparsest solution, {\it Communications
		on Pure and Applied Mathematics: A Journal Issued by the Courant Institute of
		Mathematical Sciences\/}, {\bf 59}(6), 797--829.
	
	\bibitem[Ekblom(1987)]{ekblom1987l1}
	Ekblom, H., 1987.
	\newblock The l1-estimate as limiting case of an lp-or huber-estimate, in {\em
		Statistical data analysis based on the L1-norm and related methods:
		31/08/1987-04/09/1987\/}, pp. 109--116, Elsevier.
	
	\bibitem[Farquharson(2007)]{farquharson2007constructing}
	Farquharson, C.~G., 2007.
	\newblock Constructing piecewise-constant models in multidimensional
	minimum-structure inversions, {\it Geophysics\/}, {\bf 73}(1), K1--K9.
	
	\bibitem[Farquharson \& Oldenburg(2004)]{farquharson2004comparison}
	Farquharson, C.~G. \& Oldenburg, D.~W., 2004.
	\newblock A comparison of automatic techniques for estimating the
	regularization parameter in non-linear inverse problems, {\it Geophysical
		Journal International\/}, {\bf 156}(3), 411--425.
	
	\bibitem[Goebel et~al.(2019)Goebel, Knight, \& Halkj{\ae}r]{goebel2019mapping}
	Goebel, M., Knight, R., \& Halkj{\ae}r, M., 2019.
	\newblock Mapping saltwater intrusion with an airborne electromagnetic method
	in the offshore coastal environment, monterey bay, california, {\it Journal
		of Hydrology: Regional Studies\/}, {\bf 23}, 100602.
	
	\bibitem[Guillemoteau et~al.(2022)Guillemoteau, Vignoli, Barreto, \&
	Sauvin]{guillemoteau2022sparse}
	Guillemoteau, J., Vignoli, G., Barreto, J., \& Sauvin, G., 2022.
	\newblock Sparse laterally constrained inversion of surface-wave dispersion
	curves via minimum gradient support regularization, {\it Geophysics\/}, {\bf
		87}(3), R281--R289.
	
	\bibitem[Hansen(2010)]{hansen2010discrete}
	Hansen, P.~C., 2010.
	\newblock {\it Discrete inverse problems: insight and algorithms\/}, vol.~7,
	Siam.
	
	\bibitem[Heagy et~al.(2017)Heagy, Cockett, Kang, Rosenkjaer, \&
	Oldenburg]{heagy2017framework}
	Heagy, L.~J., Cockett, R., Kang, S., Rosenkjaer, G.~K., \& Oldenburg, D.~W.,
	2017.
	\newblock A framework for simulation and inversion in electromagnetics, {\it
		Computers {\&} Geosciences\/}, {\bf 107}, 1--19.
	
	\bibitem[Hermans et~al.(2012)Hermans, Vandenbohede, Lebbe, Martin, Kemna,
	Beaujean, \& Nguyen]{hermans2012imaging}
	Hermans, T., Vandenbohede, A., Lebbe, L., Martin, R., Kemna, A., Beaujean, J.,
	\& Nguyen, F., 2012.
	\newblock Imaging artificial salt water infiltration using electrical
	resistivity tomography constrained by geostatistical data, {\it Journal of
		Hydrology\/}, {\bf 438}, 168--180.
	
	\bibitem[Hermans et~al.(2016)Hermans, Kemna, \& Nguyen]{hermans2016covariance}
	Hermans, T., Kemna, A., \& Nguyen, F., 2016.
	\newblock Covariance-constrained difference inversion of time-lapse electrical
	resistivity tomography datatl-ert covariance-constrained inversion, {\it
		Geophysics\/}, {\bf 81}(5), E311--E322.
	
	\bibitem[Hunziker et~al.(2015)Hunziker, Thorbecke, \&
	Slob]{hunziker2015electromagnetic}
	Hunziker, J., Thorbecke, J., \& Slob, E., 2015.
	\newblock The electromagnetic response in a layered vertical transverse
	isotropic medium: A new look at an old problem, {\it Geophysics\/}, {\bf
		80}(1), F1--F18.
	
	\bibitem[Jones et~al.(2001)Jones, Oliphant, Peterson, et~al.]{scipy}
	Jones, E., Oliphant, T., Peterson, P., et~al., 2001.
	\newblock {SciPy}: Open source scientific tools for {Python}, [Online; accessed
	12-01-2019].
	
	\bibitem[Kemna(2000)]{kemna2000tomographic}
	Kemna, A., 2000.
	\newblock {\it Tomographic inversion of complex resistivity: Theory and
		application\/}, Der Andere Verlag.
	
	\bibitem[Key(2009)]{key20091d}
	Key, K., 2009.
	\newblock 1d inversion of multicomponent, multifrequency marine csem data:
	Methodology and synthetic studies for resolving thin resistive layers, {\it
		Geophysics\/}, {\bf 74}(2), F9--F20.
	
	\bibitem[Klose et~al.(2022)Klose, Guillemoteau, Vignoli, \&
	Tronicke]{klose2022laterally}
	Klose, T., Guillemoteau, J., Vignoli, G., \& Tronicke, J., 2022.
	\newblock Laterally constrained inversion ({LCI}) of multi-configuration {EMI}
	data with tunable sharpness, {\it Journal of Applied Geophysics\/}, {\bf
		196}, 104519.
	
	\bibitem[Lebbe \& Pede(1986)]{lebbe1986salt}
	Lebbe, L. \& Pede, K., 1986.
	\newblock Salt-fresh water flow underneath old dunes and low polders influenced
	by pumpage and drainage in the western belgian coastal plain, in {\em Salt
		water intrusion meeting. 9\/}, pp. 199--220.
	
	\bibitem[Lee et~al.(2006)Lee, Wasilewski, Gommers, Wohlfahrt, O'Leary, \&
	Nahrstaedt]{lee2006pywavelets}
	Lee, G., Wasilewski, F., Gommers, R., Wohlfahrt, K., O'Leary, A., \&
	Nahrstaedt, H., 2006.
	\newblock Pywavelets--wavelet transforms in python.
	
	\bibitem[Linde et~al.(2015)Linde, Renard, Mukerji, \&
	Caers]{linde2015geological}
	Linde, N., Renard, P., Mukerji, T., \& Caers, J., 2015.
	\newblock Geological realism in hydrogeological and geophysical inverse
	modeling: A review, {\it Advances in Water Resources\/}, {\bf 86}, 86--101.
	
	\bibitem[Liu et~al.(2017)Liu, Farquharson, Yin, \& Baranwal]{liu2017wavelet}
	Liu, Y., Farquharson, C.~G., Yin, C., \& Baranwal, V.~C., 2017.
	\newblock Wavelet-based 3-d inversion for frequency-domain airborne em data,
	{\it Geophysical Journal International\/}, {\bf 213}(1), 1--15.
	
	\bibitem[Macnae \& Milkereit(2007)]{macnae2007developments}
	Macnae, J. \& Milkereit, B., 2007.
	\newblock Developments in broadband airborne electromagnetics in the past
	decade, in {\em Proceedings of Exploration\/}, vol.~7, pp. 387--398.
	
	\bibitem[Mallat(1999)]{mallat1999wavelet}
	Mallat, S., 1999.
	\newblock {\it A wavelet tour of signal processing\/}, Elsevier.
	
	\bibitem[Mikucki et~al.(2015)Mikucki, Auken, Tulaczyk, Virginia, Schamper,
	S{\o}rensen, Doran, Dugan, \& Foley]{mikucki2015deep}
	Mikucki, J.~A., Auken, E., Tulaczyk, S., Virginia, R., Schamper, C.,
	S{\o}rensen, K., Doran, P., Dugan, H., \& Foley, N., 2015.
	\newblock Deep groundwater and potential subsurface habitats beneath an
	antarctic dry valley, {\it Nature communications\/}, {\bf 6}(1), 1--9.
	
	\bibitem[Nittinger \& Becken(2018)]{nittinger2018compressive}
	Nittinger, C. \& Becken, M., 2018.
	\newblock Compressive sensing approach for two-dimensional magnetotelluric
	inversion using wavelet dictionaries, {\it Geophysical Prospecting\/}, {\bf
		66}(4), 664--672.
	
	\bibitem[Nittinger \& Becken(2016)]{nittinger2016inversion}
	Nittinger, C.~G. \& Becken, M., 2016.
	\newblock Inversion of magnetotelluric data in a sparse model domain, {\it
		Geophysical Journal International\/}, {\bf 206}(2), 1398--1409.
	
	\bibitem[Paasche \& Tronicke(2007)]{paasche2007cooperative}
	Paasche, H. \& Tronicke, J., 2007.
	\newblock Cooperative inversion of 2d geophysical data sets: A zonal approach
	based on fuzzy c-means cluster analysis, {\it Geophysics\/}, {\bf 72}(3),
	A35--A39.
	
	\bibitem[Pfaffhuber et~al.(2017)Pfaffhuber, Lysdahl, S{\o}rmo, Skurdal,
	Thomassen, Ansch{\"u}tz, \& Scheibz]{pfaffhuber2017delineating}
	Pfaffhuber, A.~A., Lysdahl, A.~O., S{\o}rmo, E., Skurdal, G.~H., Thomassen, T.,
	Ansch{\"u}tz, H., \& Scheibz, J., 2017.
	\newblock Delineating hazardous material without touching—aem mapping of
	norwegian alum shale, {\it First Break\/}, {\bf 35}(8).
	
	\bibitem[Podgorski et~al.(2013)Podgorski, Auken, Schamper, Vest~Christiansen,
	Kalscheuer, \& Green]{podgorski2013processing}
	Podgorski, J.~E., Auken, E., Schamper, C., Vest~Christiansen, A., Kalscheuer,
	T., \& Green, A.~G., 2013.
	\newblock Processing and inversion of commercial helicopter time-domain
	electromagnetic data for environmental assessments and geologic and
	hydrologic mapping, {\it Geophysics\/}, {\bf 78}(4), E149--E159.
	
	\bibitem[Siemon et~al.(2009)Siemon, Auken, \&
	Christiansen]{siemon2009laterally}
	Siemon, B., Auken, E., \& Christiansen, A.~V., 2009.
	\newblock Laterally constrained inversion of helicopter-borne frequency-domain
	electromagnetic data, {\it Journal of Applied Geophysics\/}, {\bf 67}(3),
	259--268.
	
	\bibitem[Siemon et~al.(2019)Siemon, van Baaren, Dabekaussen, Delsman, Dubelaar,
	Karaoulis, \& Steuer]{siemon2019automatic}
	Siemon, B., van Baaren, E., Dabekaussen, W., Delsman, J., Dubelaar, W.,
	Karaoulis, M., \& Steuer, A., 2019.
	\newblock Automatic identification of fresh--saline groundwater interfaces from
	airborne electromagnetic data in zeeland, the netherlands, {\it Near Surface
		Geophysics\/}, {\bf 17}(1), 3--25.
	
	\bibitem[Su et~al.(2021)Su, Yin, Liu, Ren, Zhang, Qiu, Xiong, \&
	Baranwal]{su2021sparse}
	Su, Y., Yin, C., Liu, Y., Ren, X., Zhang, B., Qiu, C., Xiong, B., \& Baranwal,
	V.~C., 2021.
	\newblock Sparse-promoting 3-d airborne electromagnetic inversion based on
	shearlet transform, {\it IEEE Transactions on Geoscience and Remote
		Sensing\/}.
	
	\bibitem[Thibaut et~al.(2021)Thibaut, Kremer, Royen, Ngun, Nguyen, \&
	Hermans]{thibaut2021new}
	Thibaut, R., Kremer, T., Royen, A., Ngun, B.~K., Nguyen, F., \& Hermans, T.,
	2021.
	\newblock A new workflow to incorporate prior information in minimum gradient
	support (mgs) inversion of electrical resistivity and induced polarization
	data, {\it Journal of Applied Geophysics\/}, {\bf 187}, 104286.
	
	\bibitem[Tikhonov(1943)]{tikhonov1943stability}
	Tikhonov, A.~N., 1943.
	\newblock On the stability of inverse problems, in {\em Dokl. Akad. Nauk
		SSSR\/}, vol.~39, pp. 195--198.
	
	\bibitem[Viezzoli et~al.(2008)Viezzoli, Christiansen, Auken, \&
	S{\o}rensen]{viezzoli2008quasi}
	Viezzoli, A., Christiansen, A.~V., Auken, E., \& S{\o}rensen, K., 2008.
	\newblock Quasi-3d modeling of airborne tem data by spatially constrained
	inversion, {\it Geophysics\/}, {\bf 73}(3), F105--F113.
	
	\bibitem[Vignoli et~al.(2015)Vignoli, Fiandaca, Christiansen, Kirkegaard, \&
	Auken]{vignoli2015sharp}
	Vignoli, G., Fiandaca, G., Christiansen, A.~V., Kirkegaard, C., \& Auken, E.,
	2015.
	\newblock Sharp spatially constrained inversion with applications to transient
	electromagnetic data, {\it Geophysical Prospecting\/}, {\bf 63}(1), 243--255.
	
	\bibitem[Wait(1951)]{wait1951magnetic}
	Wait, J.~R., 1951.
	\newblock The magnetic dipole over the horizontally stratified earth, {\it
		Canadian Journal of Physics\/}, {\bf 29}(6), 577--592.
	
	\bibitem[Werthmüller(2017)]{werthmueller2017open}
	Werthmüller, D., 2017.
	\newblock An open-source full 3d electromagnetic modeler for 1d {VTI} media in
	python: empymod, {\it {GEOPHYSICS}\/}, {\bf 82}(6), WB9--WB19.
	
	\bibitem[Zeuwts(1991)]{zeuwts1991hydrogeologie}
	Zeuwts, L., 1991.
	\newblock Hydrogeologie en hydrochemie van de ijzervlakte tussen de
	frans-belgische grens en avekapelle-pervijze (westelijke kustvlakte), {\it
		Doctoraatsproefschrift, Universiteit Gent. Boorbeschrijving: 0, 00-0\/}, {\bf
		25}, 25--0.
	
	\bibitem[Zhu et~al.(1997)Zhu, Byrd, Lu, \& Nocedal]{zhu1997algorithm}
	Zhu, C., Byrd, R.~H., Lu, P., \& Nocedal, J., 1997.
	\newblock Algorithm 778: L-bfgs-b: Fortran subroutines for large-scale
	bound-constrained optimization, {\it ACM Transactions on Mathematical
		Software (TOMS)\/}, {\bf 23}(4), 550--560.
	
\end{thebibliography}

\appendix
\section{$\beta$-cooling strategy}
\label{ap:betareducing}
To reduce the computational burden, the discrepancy principle is often combined with
	an imposed cooling-schedule-type behaviour on the regularization parameter $\beta$ \citep{farquharson2004comparison}. The cooling is usually done based on the minimum change in the misfit. Here, we have adopted different cooling or $\beta$-reducing rules. We use a warm-start strategy. Starting from a relatively high initial regularization parameter $\beta_0$, we sequentially run a new set of iterations with smaller regularization parameters, with the inversion model from the previous set of iterations as starting model. For each new set of iterations, we let the optimalization algorithm converge (for specific convergence details, see \citet{deleersnyder2021inversion}) and count the number of iterations that were required. Then, the regularization parameter $\beta$ is reduced, depending on the number of iterations of the previous set of iterations, as few iterations suggest that the objective function has not fundamentally changed with the new reduced $\beta$-value. If the number of iterations of the previous is more than 50, the regularization parameter is reduced with only 10\%. When the number of iterations is between 20 and 50, the regularization parameter is reduced with 25\%. When the number of iterations is below 20, the regularization parameter is reduced with 40\%. These reducing factors are being decided through experience. Loosening the conditions (fewer iterations, higher reducing factors) will reduce the overall computational cost, but increases the risk of overshooting the target data misfit. Adopting stricter conditions (higher number of iterations, lower reducing factors) will increase the computation time. It remains possible, yet less convenient, to restart from the set of iterations $j+1$ (which are stored) and redefine the reducing schedule, when appropriate.

\section{Building block metaphor underlying the intuition of the scale dependency}
\label{ap:buildingblock}
The idea behind the scale-dependency can be understood with the building block metaphor. An inversion model is built with all the (compactly supported) building blocks with different widths (or dilatation parameter $n$). When playing with wavelet-building blocks, there is a rule that the blocks can only be placed at specific locations, determined by the parameter $k$. This is a sparsity-based regularization scheme, thus one tries to build the inversion model with the least number of building blocks, or with as much vanishing wavelet coefficients as possible. Adding the scale-dependency to the scheme is basically adding the rule that for using smaller or narrower building blocks (with high dilatation parameter $n$), a higher price has to be paid. This is intuitively clear because a minimum structure model is simple if it gives priority to larger structures, rather than to small details. Using multiple small building blocks usually corresponds to adding (potentially too many) details to the model. The idea of adding scale-dependency to the regularization scheme also has a theoretical origin, studied in \citet{deleersnyder2021inversion}.


\label{lastpage}

\end{document}